\documentclass{aa}  

\usepackage{subfig}
\usepackage{stfloats}
\usepackage{graphicx}
\usepackage[normalem]{ulem}
\bibpunct{(}{)}{;}{a}{}{,}
\usepackage{txfonts}
\usepackage[table,xcdraw]{xcolor}
\usepackage{amssymb,amsmath,latexsym,mathrsfs}
\usepackage{color}
\usepackage[hidelinks]{hyperref}

\usepackage{textcomp}
\usepackage{gensymb}
\usepackage{soul}

\begin{document}

\title{ Exploring Jet Structure and Dynamics in Short Gamma Ray Bursts: A Case Study on GRB
090510
}
\titlerunning{Exploring jet structure and dynamics in SGRBs}
\authorrunning{Saji et al.}

\author{J. Saji\inst{1}\fnmsep\thanks{E-mail: jsaji@cft.edu.pl},                       M.G. Dainotti\inst{2,3,4,5}\fnmsep\thanks{E-mail: maria.dainotti@nao.ac.jp},  S. Bhardwaj\inst{2,3}\fnmsep\thanks{E-mail: shubham.bhardwaj@grad.nao.ac.jp},
        \and A. Janiuk\inst{1}\fnmsep\thanks{E-mail: agnieszka.janiuk@cft.edu.pl}
        }

\institute{Center for Theoretical Physics, Polish Academy of Sciences, Al. Lotników 32/46, 02-668 Warsaw, Poland
           \and Division of Science, National Astronomical Observatory of Japan, 2-21-1 Osawa, Mitaka, Tokyo 181-8588, Japan
           \and Department of Astronomical Sciences, The Graduate University for Advanced Studies, SOKENDAI, Shonankokusaimura, Hayama, Miura District, Kanagawa 240-0115, Japan
           \and Space Science Institute, 4765 Walnut St Ste B, Boulder, CO 80301, USA
           \and Nevada Center for Astrophysics, University of Nevada, 4505 Maryland Parkway, Las Vegas, NV 89154, USA}

   \date{}

\abstract
{Gamma-ray bursts observed in high-energies allow the investigation of the emission processes of these still puzzling events. \\
 In this study, we perform general relativistic magnetohydrodynamic (GRMHD) simulations to investigate GRB 090510, a peculiar short GRB detected by \textit{Fermi}-LAT. Our primary goal is to model the energetics, jet structure, variability, and opening angle of the burst to understand its underlying physical conditions.
We tested the 2D and 3D models and estimated the time scale of variability. The predicted energetics and the jet opening angle reconcile with the observed ones with 1 $\sigma$ when considering that the jet opening angles also evolve with redshift.
Furthermore, we extend our analysis by incorporating dynamical ejecta into selected models to study its impact on jet collimation at smaller distances. In addition, we investigated a suite of models exhibiting a broad range of observable GRB properties, thereby extending our understanding beyond this specific event.\\
}

 {}


\maketitle
%
 
\section{Introduction}
Gamma-ray bursts (GRBs) are intense pulses of high-energy radiation emitted by ultra-relativistic jets aligned with our line of sight as observed from Earth \citep{2015PhR...561....1K, 10.1088/2514-3433/aae164}. The observed bursts fall into two categories based on their $T_{90}$ duration \citep{1993ApJ...413L.101K}: short GRBs lasting less than 2 seconds are generally attributed to the mergers of compact binaries \citep{Paczynski1986}, while long GRBs ($T_{90} < 2\,\text{s}$) are thought to be driven by the collapse of massive stars \citep{MacFadyen1999}. 
In both cases, a fast jet is emitted along the symmetry axis of the central engine. 
 Several mechanisms have been proposed for powering these relativistic jets, including the Blandford–Znajek (BZ) process from a rotating black hole \citep{1977MNRAS.179..433B}, magnetised neutrino-driven winds \citep{Dessart_2009, 10.1046/j.1365-2966.2003.07032.x}, and spin-down of a magnetar \citep{10.1111/j.1365-2966.2011.18280.x, doi:10.1142/S021827181842004X}.

These relativistic jets are collimated, and the collimation degree is characterised by a half jet opening angle, $\theta_{j}$. However, estimating the $\theta_{j}$ poses a significant challenge, as it demands meticulous observations and a multi-wavelength analysis, which is often hindered due to the scarcity of observations or the limited availability of observation programs and telescope time. Moreover, these observed limitations impede the study of $\theta_{j}$ and properties like jet variability, adding more complexity to our understanding of jet dynamical properties. Various methods exist for determining $\theta_{j}$ from observational data \citep{
Frail_2001, Bloom_2003, 2006Natur.442.1014S,Pian2006Natur.442.1011P, 2006A&A...454L.123W, Ghirlanda_2007, Berger2014ARA&A..52...43B, Fong_2015, Pescalli2015MNRAS.447.1911P,Troja2016ApJ...827..102T, 2016ApJ...818...18G, Lloyd2019MNRAS.488.5823L,LR20,Fong2021ApJ...906..127F}. However, it is essential to note that reaching a consensus on a precise value from various methods is challenging, due to the model dependence on microphysical parameters and other scaling assumptions.

In this context, we employ a physical model of jet launching from a GRB central engine to enhance our understanding of the dynamical properties of GRB jets, focusing on $\theta_{j}$, variability, and energetics. 
We propose to model the dynamical evolution of the central engine, produced by the two neutron star (NS) mergers, via high-resolution 2D and 3D general relativistic magnetohydrodynamic (GRMHD) simulations, focusing on the estimate of $\theta_{j}$, their variability, energy, and luminosities. 
Furthermore, we synergise with existing methods to infer the $\theta_{j}$ \citep{2012ApJ...745..168L, 2014AnP...526..340A, Fong_2015, Pescalli2015MNRAS.447.1911P, 2016ApJ...818...18G} and variability analysis \citep{2012ApJ...744..141B, 2013MNRAS.432..857M, 2015ApJ...811...93G} from observations. 
The final aim of the paper is to compare the observational properties of this GRB either derived in the literature or by us with the parameters derived by the modelling.

The paper is structured as follows: Sec. \ref{Data analysis} discusses the observational properties of GRB 090510, including the calculation of the bolometric luminosity, isotropic energy, minimum variability timescale calculation, and the jet opening angle determination. Sec. \ref{numerical simulations} presents the parameters used in our numerical simulations 
and highlights the similarities and differences between the 2D and 3D models. 
Sec. \ref{sec:results} shows the results from our numerical simulation. In Sec. \ref{Discussion and conclusions}, we discuss our findings. Finally, in Sec. \ref{sec:summary and conclusions}, we summarize our work and draw conclusions.

\section{GRB 090510}\label{Data analysis}
We chose for our comparison the short GRB 090510. 
This GRB possesses the plateau emission among the GRBs observed by the \textit{Fermi}-Large Area Telescope \citep[LAT;][]{2009ApJ...697.1071A} from August 2008 until August 2016 \citep{Dainotti2021ApJS..255...13D} with observed redshifts given in the Second \textit{Fermi}-LAT GRB Catalog \citep[2FLGC;][]{Ajello_2019}. To select this GRB, we first considered 19 GRBs analyzed in the 2FLGC from July 2008 until May 2016 following the analysis performed in \cite{Dainotti2021ApJS..255...13D}  that were identified as the brightest sources based on maximum likelihood analysis, with a Test Statistics\footnote{the TS is defined as twice the logarithm of the ratio of the maximum likelihood obtained using a model including the GRB over the maximum likelihood value of the model that excludes the GRB}, $TS>64$. Out of these 19 GRBs, we select those that can be fitted with a broken power law (BPL) in the 2FLGC and have reliable fitting parameters for which the error bars do not exceed the values of the best-fit parameters themselves. This reduces the sample to only three GRBs: GRB 090510, 090902B, and 160509A, with the only short GRB being 090510, making it a particularly interesting case for further investigation.  GRB 090510 benefits from comprehensive multi-wavelength coverage from Fermi-GBM/LAT and Swift-XRT, with well-constrained redshift, bolometric energy estimates, and variability measurements. These characteristics make it particularly well-suited as a benchmark case for our GRMHD jet modeling.

The observational properties of GRB 090510 are summarised in Table \ref{tab:parameters}. We simulate jet models in which the central engine parameters are calibrated to reproduce key observables of this GRB, including its energetics, luminosity, and jet opening angle.

\begin{center}
  \begin{table}[h!]
  \centering
  \begin{tabular}{ c c }
    \hline
     \multicolumn{2}{c}{090510} \\
\hline
\hline
$T_{\rm{GBM,90}}$ (s) & 0.6 \\
$E_{\rm{iso}}$ (erg) & $(9.97 \pm 0.51) \times 10^{52}$ \\
redshift ($z$) & 0.903 \\
distance (Gpc) & 5.86 \\

\hline
\hline
  \end{tabular}
\caption{Summary of parameters for GRB 090510. $T_{\rm{GBM,90}}$ is the time in which 90\% of the total emission of the prompt is released. We also give rest-frame isotropic $E_{\rm{iso}}$, the redshift ($z$), and the corresponding distance in Gigaparsec (Gpc).}
 \label{tab:parameters}
\end{table}
\end{center}

\subsection{Luminosity and Isotropic Energy}
\label{sec:luminosity_calculation}

Since the luminosity obtained from the theoretical model is bolometric, we compute the observational counterpart by combining the high-energy (gamma-ray) and X-ray luminosities derived from multi-wavelength data.

The luminosity is computed in the following way for the total energy band:
\begin{equation}
L(E_{min}, E_{max}, t)= 4 \pi D_L^2(z, \Omega_M, h) \, F (t) \cdot K(E_{min}, E_{max}),
\label{eq: lx}
\end{equation}

\noindent
where $D_L(z, \Omega_M, h)$ is the luminosity distance computed in the flat $\Lambda$CDM cosmological model with $\Omega_M = 0.291$ and $h=0.70$ in units of $100$ $km$ $s^{-1}$ $Mpc^{-1}$, $F$ is the measured energy flux, $E_{min}, E_{max}$ are the appropriate energy bandpass for each instrument we use. The \textit{K} is the \textit{K}-correction for the cosmic expansion \citep{Bloom2001}:

\begin{equation}
K=\frac{\int_{E_{min}/(1 + z)}^{E_{max}/(1 + z)}{\Phi (E)
dE}}{\int_{E_{min}}^{E_{max}}{\Phi (E) dE}},
\label{eq: kcorrection}
\end{equation}

\noindent
where the energy spectrum $\Phi (E)$ is described by the Band function or a simple power law (PL).

The bolometric luminosity is \( L_{\rm bol} = L_{\rm GBM,LAT} + L_{\rm XRT} \). The optical contribution, being orders of magnitude lower in luminosity, has a negligible impact on the total energy budget and is thus excluded from the estimate.
Multiple studies have conducted the spectral analysis of GRB 090510. For our analysis, we adopt the fitting parameters from the optimal model from the combined \textit{Fermi}-Gamma-ray Burst Monitor \citep[GBM;][]{2009ApJ...702..791M} and LAT spectral analysis performed by the \textit{Fermi}-GBM/LAT collaboration\citep{Ackermann_2010}. 
This best-fit model, applicable during the interval 0.5 - 1s, which coincides with the peak high-energy counts in the GBM and LAT light curves (LCs), consists of a Band function with a PL component.
The afterglow peak luminosity from the \textit{Swift}-XRT analysis in the energy range 0.3-10 keV is taken from \cite{2016ApJ...829....7L}. 
This gives the bolometric luminosity of  $L_{\rm bol}$=$(3.90\pm 0.55)\times10^{53}$ erg s$^{-1}$ for GRB 090510.

For calculating $E_{\rm iso}$, we use the relation:
\begin{equation}
E_{\mathrm{iso}} = 4\pi D_L^2 S_{\mathrm{bolo}}K(1+z)^{-1} 
\label{eq: Eiso-eqn}
\end{equation}

\noindent

 We adopted the bolometric fluence, $S_{\mathrm{bolo}} = (5.03 \pm 0.25) \times 10^{-5} \, \mathrm{erg \, cm^{-2}}$, as reported by \cite{Ackermann_2010} for the 10 KeV to 30 GeV energy range.
This gives a bolometric $E_{\rm{iso}}$ of the jet to be ($9.97 \pm 0.51)\times 10^{52} \, \mathrm{erg}$. 
Both luminosities and energies computed and reported here are used for our comparison with the theoretical simulations.

\subsection{Minimum Variability Timescale}\label{MTS-obser}
The observed GRB LCs exhibit strong temporal variability. The minimum timescale variability (MTS) refers to the shortest duration within which a significant change in the count rate occurs in the observed LC. 

The measure of variability has been a longstanding and debated issue in the literature since its inception \citep{Sari_1997}, and although the variability as an intrinsic property of GRB should have a unique measure, often in the literature, values are discrepant. It is challenging to assess which of the methods is the most reliable. Below, we summarise and discuss two of the most used methods: the Bayesian Block (BB) and the wavelet analysis, and we check the value of the variability pertinent to GRB 090510.

The BB method detailed in \cite{Scargle_2013} divides the LC into various time intervals of different time widths, closely following true underlying variation in the emission. 

This algorithm is a nonparametric modelling that employs optimal segmentation analysis on sequential data, with sampling that can be arbitrary in the presence of gaps in the data and different exposure times. In the BB algorithm, each block (or bin in our case) is consistent with a probability distribution function (PDF), and the entire dataset is represented by this collection of finite PDFs.
The number of blocks and the edges of the blocks are set via the optimisation of a `fitness function', namely a goodness-of-fit statistic dependent only on the input data and the regularisation parameter. The set of blocks presents no gaps, and there is no overlapping between one block and the other, where the first and last block edges are defined by the first and last data points, respectively. 
The BB method is based on the additivity of the fitness function, and thus, the fitness of a given set of blocks is equal to the sum of the  fitness of the individual blocks. The total fitness, $F_{tot}$ for a given dataset is :
\begin{equation}
F_{tot}=\sum_{i=1}^{N}{f(B_i)}
\end{equation}
where $f(B_i)$ if the fit for an individual block and $N$ is the total number of blocks. The MTS is then obtained by identifying the shortest block width produced by the segmentation. For analysis, counts rate LC of GRB090510 from the NaI detector of \textit{Fermi}-GBM in the energy range 10 KeV to 250 KeV is used. The LCs are produced from the GRB data provided by The \textit{Fermi} Science Support Center (FSSC) and processed using \textit{Fermitools}. The MTS has been calculated to be 32 milliseconds using the BB method.

Complementing the BB approach, wavelet analysis provides an alternative and widely adopted method in GRB variability studies, particularly effective in isolating transient features across multiple frequencies. Unlike Fourier analysis, wavelet analysis employs basis functions that are localized in both time and frequency domains, enabling precise identification of variability timescales. Using wavelet analysis, \cite{2013MNRAS.432..857M} found an MTS of $4.9^{+1.1}_{-0.9}$ milliseconds for GRB 090510.  We will be exploring how these observational methods compare with results from numerical simulations in later sections of this paper.%

\subsection{Jet Opening Angle From Observation}\label{sec:jet-opening}
As discussed in the introduction, the determination of the $\theta_{j}$ is challenging and several ways have been proposed in the literature using various observational and theoretical methods \citep{2009ApJ...706L..33G, Ackermann_2010, 2010ApJ...720.1008C, 2011ApJ...739...47F, 2011ApJ...733...22H, 2011MNRAS.414.1379P, 2012A&A...548A.101N, 2014ApJ...787L..32E, 2016ApJ...831...22F, 2016ApJ...818...18G}. These studies report values that are typically less than \(1^\circ\) but with significant uncertainties among the several measurements, often differing by an order of magnitude across different models and approaches. This wide range reflects the inherent uncertainties in deriving \(\theta_j\) from observational data, which stem from differences in jet structure assumptions, energy distribution models, assumption of the density of the circumburst medium, and observational biases \citep{Sari1999, Chevalier_2000}.

To simplify calculating $\theta_{j}$, the luminosity and energy output of GRBs are typically computed under the assumption of isotropic emission. 
Despite several methods available to estimate $\theta_{j}$ \citep{2004ApJ...616..331G, 2012ApJ...745..168L, Fong_2015, 2016ApJ...818...18G}, not all GRBs have this parameter available.  In this work, we use an indirect derivation of $\theta_{j}$ based on the method detailed in \cite{Pescalli2015MNRAS.447.1911P} where the $E_{\rm{peak}} - E_{\rm{\gamma}}$ relation \citep{2004ApJ...616..331G} and the $E_{\rm{peak}} - E_{\rm{iso}}$ relation \citep{2002A&A...390...81A, 2009A&A...508..173A} have been used and considered reliable standardizable candles.

From \cite{2004ApJ...616..331G},
we have:
\begin{equation}
   E_{\rm{peak}} = A_G \times \Biggl(\frac{E_{\rm{\gamma} \rm{, erg}}}{4.3 \times 10^{50} erg}\Biggl)^{k_{G}}
   \label{Ghirlanda}
\end{equation} 
\noindent
where $E_{\rm{\gamma} \rm{, erg}} = E_{\rm{iso}} (1 - \cos\theta_{j})$, 
$A_G=267 $ and $k_G=0.706 \pm 0.047$ are the normalisation and the slopes of the Ghirlanda relation, respectively.

From \cite{2002A&A...390...81A},
we have:
\begin{equation}
   E_{\rm{peak}} = A_A \times \Biggl(\frac{E_{\rm{iso}}}{10^{52} \rm{erg}}\Biggl)^{k_A}
   \label{Amati}
\end{equation}
\noindent
where $A_A=100$ and $k_A=0.52\pm 0.06$ are the normalisation and the slope of the Amati relation, respectively.

From equaling these two equations, we obtain:
\begin{equation}
\cos\theta_{j} = 1 - \Bigl(4.3 \times 10^{50}\Bigl) \times\, \Biggl(\frac{A_A}{A_G}\Biggl)^{\frac{1}{k_G}} \times \, E_{\rm{iso}}^{{\frac{k_A}{k_G} - 1}} \times \Bigl(10^{52}\Bigl)^{\frac{k_A}{k_G}}.
\end{equation}

Substituting constants and performing error propagation, we obtain:

\begin{equation}
    \cos\theta_{j} = 1 - \frac{(0.249 \pm 0.023) \times (4.3 \times 10^{50})}{10^{38.324 \pm 5.096} \times E_{\rm{iso}}^{0.263 \pm 0.098}}
\label{Eq:jet-angle-1}
\end{equation}

\noindent
which can further be approximated as:

\begin{equation}
  \cos\theta_{j} \approx  1 - \frac{5.078 \times 10^{11}}{E_{\rm{iso}}^{0.263}}
\label{Eq:jet-angle}
\end{equation}

Thus, for our GRB, since we have the measure of the isotropic energy ($E_{\rm{iso}}$), the angle $\theta_{j}$ can be calculated and it is $\theta_{j}\approx 6^\circ$\footnote{ When we apply the differential error propagation to derive the uncertainties in $\theta_{j}$ based on Eq. \ref{Eq:jet-angle-1}, namely on $E_{iso}$ and on $k_A$ and $k_G$ we obtain a very large uncertainty of $\pm$32$^\circ$ without redshift evolution correction. This source of uncertainties is large and, thus, can lead to nonphysical values. In our comparison with the simulations, we will refer to an ideal case of the Amati and Ghirlanda relation, which carries no uncertainties.}. We note that independent observational methods have reported smaller values of \(\theta_j\) for this GRB, highlighting the methodological dispersion discussed above. Our approach provides a consistent estimate based on isotropic energy, but should be calibrated across a broader GRB sample as more high-quality spectral analyses become available.

However, this value of $\theta_{j}$ needs to be corrected for the evolution of redshift ($z$). Following \cite{Lloyd2019MNRAS.488.5823L}, we correct for the highly statistically significant ($\approx 5 \sigma$) anti-correlation between $\theta_{j}$ and $z$, with a functional form that reads as follows: 
$\theta_{j} \sim (1 + z)^{-0.75 \pm 0.2}$. Using Eq. \ref{Eq:jet-angle} and from the evaluation of the observational $E_{\rm{iso}} = (9.97 \pm 0.51) \times 10^{52}$ erg (Sec. \ref{sec:luminosity_calculation}), and $z$ = 0.903, we obtain $\theta_{j}$.

We notice, that the value of $\theta_{j}$ needs to be corrected for the evolution with redshift ($z$). 
This relation is not a consequence of observational selection bias alone, but rather reflects an intrinsic redshift evolution in jet opening angles derived using the Efron–Petrosian method \citep{Efron1992}, which accounts for such biases. Therefore, this correction is distinct from the redshift dependence already present through the E$_{\rm iso}$ in Eq. \ref{Eq:jet-angle}. 
We obtain $\theta_{j} \approx 10^\circ$ after applying the redshift evolution correction.

\section{Numerical Simulations of GRB Engines}\label{numerical simulations}

Our model of the SGRB engine adopts a scenario of the binary NS merger progenitor but starts when the post-merger compact object has already been formed. The typical outcome of such scenario  is the short-living stage of a hyper-massive NS \citep{Sekiguchi2011PhRvL.107e1102S, Hotokezaka2013PhRvD..88d4026H}, which immediately collapses into BH.  While a BH is not the only plausible central engine, GRB090510 - our target GRB in this work has been previously studied and argued to be powered by a BH formed after an NS-NS merger \citep{Ruffini_2016}.

\subsection{Code}
We use the GRMHD code HARM \citep{Gammie_2003, Noble_2006, Sapountzis_2019} to perform a suite of numerical simulations of the accretion flow around a Kerr BH. HARM code employs a finite volume, shock-capturing scheme to solve the hyperbolic system of partial differential equations in their conservative form. It uses the Harten, Lax, van Leer (HLL) scheme to numerically compute the corresponding flux function.

The code follows the evolution of the gas and magnetic field by solving the continuity, energy-momentum conservation, and induction equations:
\begin{equation}
    \nabla_\mu(n u^{\mu})= 0,
\end{equation}    
\begin{equation}
    \nabla_\mu(T^{\mu\nu}) = 0,
\end{equation}
\begin{equation}
    \nabla_\mu(u^\nu b^\mu - u^\mu b^\nu) = 0,
\end{equation}
where $n$ is the baryon number density in the fluid frame, $u^{\mu}$ is the four-velocity of the gas, and $b^{\mu}$ is the magnetic four-vector. 

The stress-energy tensor consists of matter and electromagnetic parts: $T^{\mu \nu} = T^{\mu \nu}_{\mathrm{gas}} + T^{\mu \nu}_{\mathrm{EM}}$, where:

\begin{equation}\label{eqn:T_MA}
    T^{\mu \nu}_{\mathrm{gas}} = (\rho + u + p )u^{\mu}u^{\nu} + pg^{\mu \nu},
\end{equation}

and

\begin{equation}\label{eqn:T_EMb}
    T^{\mu \nu}_{\mathrm{EM}} = b^2 u^{\mu}u^{\nu} + \frac{1}{2}b^2 g^{\mu \nu} - b^{\mu} b^{\nu},
\end{equation}

Here $u$ is the internal energy, and $p$ is the gas pressure.
The closing equation is that of pressure relation with density, which defines the equation of state (EoS). In the current simulations, we use an adiabatic form, $p=(\gamma_{ad}-1)u$, with $\gamma_{ad}=4/3$.

We adopt dimensionless units, with $G = c = M = 1$, for our simulations. Thus, the length in the code units is given by $r_g$ = $GM/c^2$, and the time is given by $t_g$ = $GM/c^3$, where $M$ is the BH mass. 

The BH mass is fixed in all our models to be 3$M\odot$.
As we have to compare the results with GRB 090510, all main models were run for a duration of $\sim50000 t_g$, effectively covering 0.6 s in real-time, which matches the \textit{Fermi}-GBM estimated $T_{GBM, 90}$ for GRB 090510. We restrict the simulation duration to the prompt gamma-ray emission phase, as the modelled jet dynamics are not intended to capture the later afterglow evolution.

\subsection{Model}

The accretion disk is modelled based on \cite{1976ApJ...207..962F}, hereafter FM, solution for a steady-state, pressure-supported fluid in a Kerr BH's potential. The torus solution is parameterised by two values, the radius of its pressure maximum, $r_{max}$, and the radius of the inner cusp location, $r_{in}$, and it is constructed for a given BH spin, $a$.

The FM torus is embedded in a poloidal magnetic field. The field geometry resembles that of a circular wire field configuration, and the non-vanishing component of magnetic vector potential is given by:

\begin{center}
\begin{equation}
    A_{\phi}(r, \theta) = A_0 \frac{(2-k^2)K(k^2) - 2E(k^2)}{k \sqrt{4Rr \sin \theta}} \label{eq-magfield}
\end{equation}

\[
    k = \sqrt{\frac{4Rr \sin \theta}{R^2 + r^2 + 2Rr \sin \theta}}
\]
\end{center}
\noindent
Here, $E$ and $K$ are complete elliptic functions, and $A_0$ is the field normalisation constant \citep{Sapountzis_2019}. The radius of the circular wire is taken to be the $r_{max}$ of the FM torus. This magnetic field is scaled across the torus using the plasma-beta factor given by the ratio of initial gas pressure to magnetic pressure, $\beta = p/p_{mag}$, with $p_{mag}=b^{2}/2$. The total mass of the torus is calculated using a physical density scaling in the cgs units. To compute this value, we use an arbitrary mass unit, $M_{unit} = 1.5 \times 10^{-5} M_{\odot}$, and hence the torus mass is given by $M_{torus}= M_{unit} \int \rho(r,\theta,\phi) \sqrt{-g} dr\,d\theta\,d\phi $, where density $\rho$ is expressed in code units. 

The FM accretion disk in our model accretes matter into the Kerr BH due to magneto-rotational instability (MRI). Magnetised tori launch relativistic jets along the rotation axis of the Kerr BH through the Blandford-Znajek mechanism \citep[BZ;][]{1977MNRAS.179..433B}. We have employed different models with varying parameters, including plasma, $\beta$ parameter, BH spin, and disk mass (see Table \ref{tab:Sim_Models} for model parameters). These variations have provided us with a number of unique jet signatures to be compared with the observed properties. As this study did not incorporate a radiative transport mechanism to explicitly connect simulations to spectral and temporal LC characteristics, we focus on properties like total energy, $\theta_{j}$, variability timescales, and $\Gamma$ factor. These are interconnected physical properties that are key to understanding the observed phenomena.

\subsubsection{Dynamical Ejecta}\label{Dynamic_ejecta_1}

We have extended the parameter space to include two models with dynamical ejecta setups, which depict GRB environments more realistically. This enables the exploration of ejecta effects on jet structure, collimation, and energetics.

There are multiple parameters that shape a GRB jet. While the accretion disk winds \citep{10.1093/mnras/stab2982,10.1093/mnras/stab723,galaxies10050093} are one factor, another important parameter is the dynamical ejecta (DE). During the inspiral phase of the binary NS (BNS) merger, tidal interactions and shocks can lead to the ejection of a considerable amount of mass. This neutron-rich ejected material presents a potential site for r-process nucleosynthesis, which may result in observable phenomena such as kilonova emissions \citep{2017ApJ...848L..24V, Metzger_2017, 2018ApJ...860...62G, 2020NatAs...4...77J, rastinejad2024}. 

DE and accretion disk winds interact with the jet by exerting external pressure and modifying the surrounding medium's density and pressure gradients, which is effectively squeezing and collimating the jet into a narrower opening angle. Since we are dealing with the narrow jet of 090510, the implementation of dynamical ejecta in the simulation setup could be a viable method to collimate our jet as well as study its effect in jet dynamics in general. 

However, estimating the mass and distribution of DE is challenging, particularly in cases like GRB 090510, where direct kilonova observations are lacking. This led us to follow the theoretical works as well as merger simulations \citep{refId0, PhysRevD.87.024001, Radice_2018, PhysRevD.110.024003} to try a range of DE configurations that can help us shape a jet as we find from observations. 
Several works have implemented ejecta configurations in GRB jet studies \citep{Bauswein_2013, Nagakura_2014, Nedora_2021, 10.1093/mnras/stab1810, 2022ApJ...933L...2G}. Given that the properties and structure of DE heavily depend on specific merger characteristics such as the mass ratio, NS spins, and their EoS, we implemented a simplified ejecta profile in our model.
The former studies suggest that the mass of DE is expected to be in the range of $10^{-4} \lesssim M_{\text{ej}} \lesssim 10^{-2}\, M_{\odot}$ and has a structure in which the density is concentrated in the equatorial plane. A convenient functional form is proposed by \cite{2022ApJ...933L...2G}:

\begin{equation}
    \rho(r_{1} < r < r_{out}, \theta) = \rho_0 r^{-\alpha} (0.1 + \sin^2 \theta)^\delta.
\end{equation}

\noindent
Here, $r_1$ represents the inner boundary of the ejecta expansion, determined by their initial velocity and the delay time between the merger and BH formation. The other parameters $\rho_0$, $\alpha$ and $\delta$, are treated as free variables, with values selected based on the general trends observed in the simulations mentioned above.
For all our models, we have used a poloidal magnetic field configuration, except for the case with DE, where even though the magnetic field is poloidal, it is restricted in radius such that the DE is kept unmagnetised. For all our models, $\alpha$ and $\delta$ were 2 and 1, respectively, and density scaling factor $\rho_{0}$ was adjusted to achieve the desired total ejecta mass in each case.

Models listed in Table \ref{tab:Sim_Models} have parameters which could reflect an SGRB scenario. 
All models, except one, are 2D (axisymmetric). The models with DE have an outer radius set at 3500$r_g$, while for the normal models, it is 1000$r_g$. Although the chosen outer radius may not fully capture the complete influence of DE on jet dynamics, but it can give us a preliminary understanding of jet-ejecta interactions close to jet base.

\subsubsection{The jet  characteristics}\label{sec:jet_characteristics}

Since the jets are Poynting-dominated, we introduce two parameters: jet energetics parameter $\mu$ and jet magnetisation parameter $\sigma$, which we will be using to quantify the properties of the simulated jet. 
The $\mu$ and $\sigma$ are defined as:

\begin{equation}
\mu = -\frac{T^r_t}{\rho u^r} \;\;\;\;\;\;\;\;  \sigma = \frac{(T_{\text{EM}})^r_t}{(T_{\text{gas}})^r_t}
\end{equation}
\noindent
where $\mu$ represents the total specific energy of the jet; the ratio of the total energy flux to the mass flux. $T^r_t$ is the radial component of the energy-momentum tensor, representing the energy flux in the radial direction, $\rho$ is the rest-mass density, and $u^r$ is the radial component of the four-velocity.
The parameter $\sigma$ is the magnetisation parameter of the jet. It is defined as the ratio of the electromagnetic energy flux to the gas energy flux.

The upper limit of the terminal $\Gamma$ factor can be obtained from the local flow quantities. The maximum terminal $\Gamma$ ($\Gamma_{\infty}$) can be approximated with the jet energetics parameter $\mu$. It is the sum of the inertial thermal energy of the plasma and its Poynting flux, which can be transferred to the bulk kinetic energy of the jets at large distances. \citep{Vlahakis_2003, Janiuk_2021, James_2022}. Since $\mu$ is a direct estimate of the total energy confined in the jet, we also utilise $\mu$ to constrain the jet region and obtain $\theta_{j}$. The opening angle is calculated by determining the $\theta$ value that encompasses 75\% of the total jet energy. We analyse the $\theta$ profile of $\mu$ at $R_{out}$ for each model.  

As for the luminosity of the jet, computed as the BZ luminosity, we calculate it at the given time snapshot as equal to $L_{BZ}=E_{unit} \int -T^{r}_{t} \sqrt{-g} d\theta d\phi$ at the BH horizon. Here, the physical scaling is given by the energy unit, $E_{unit}=M_{unit}\, c^{2}/t_{g}$.

\subsubsection{Jet Efficiency}

The efficiency ($\eta$) of converting jet energy into observable radiation is a key parameter in GRB physics but remains highly uncertain. It depends on factors such as jet dynamics, emission mechanisms, and the energy dissipation processes. Observations and theoretical studies suggest a broad range of values, reflecting the complexity of GRB jets.

Early studies, such as \cite{1997ApJ...490...92K}, demonstrated that efficiency increases with a broader range of $\Gamma$ factors among jet shells, as this enhances internal collisions and energy dissipation. Narrow $\Gamma$ factor distributions result in efficiencies as low as $\sim 2\%$, while broader distributions can raise efficiency to $\sim 40\%$. Similarly, \cite{Kobayashi_2001} showed that efficiency could reach $\sim 60\%$ for jets with non-uniform $\Gamma$ factor distributions and comparable shell masses.

Observational analyses further highlight the variability in jet efficiency. For instance, \cite{Lloyd-Ronning_2004} estimated efficiencies between $10\%$ and $99\%$ using X-ray afterglow data, while \cite{10.1111/j.1365-2966.2006.10280.x} revised this range to $1\%$-$89\%$ with updated methods. Recent studies using \textit{Fermi}-LAT and \textit{Swift}-XRT data, such as \cite{2015MNRAS.454.1073B, 2016MNRAS.461...51B}, reported efficiencies ranging from $1\%$ to $98\%$, depending on the observational wavelength and cooling effects included.

As discussed above in detail, $\eta$ of the jet produced via the BZ mechanism can span a broad range, varying from as low as 1\% to as high as 100\%. 
Given that our simulations do not account for radiative transport, it is challenging to quantify the fraction of the jet's total energy that is converted into observable radiation. Therefore, we can agree upon an estimate based on theoretical predictions regarding $\eta$ of BZ jets. 
Furthermore, \citet{2020MNRAS.491.3192H} demonstrated that assuming a $\eta$ of 10\% instead of 50\% did not alter their conclusions about the jet structure and dynamics, reinforcing the validity of our adopted value.
Hence, in this study, we adopt a radiative efficiency of 1-10\%, consistent with the approach used by \citet{10.1093/mnras/stz390}.

\section{Results of Numerical Modeling}
\label{sec:results}

Our numerical models are designed to explore a wide range of jet structures and examine different configurations. The aim is to reproduce the key observational properties of GRB 090510. Since jet structures can vary significantly depending on the properties of the accretion disk, BH, and magnetic field, our models include a diverse range of these parameters to ensure comprehensive coverage.

For GRB 090510, we have calculated the isotropic-equivalent energy, \(E_{\rm iso} = 9.97 \times 10^{52} \, \mathrm{erg}\). Using the estimated jet opening angle, \(\theta_{j} = 10^\circ\), we compute the collimation-corrected GRB energy using the relation:

\begin{equation}
E_{\rm GRB} = (1 - \cos\theta_{j}) E_{\rm iso}.
\label{eq:ejet}
\end{equation}

Substituting the values, we obtain \(E_{\rm GRB} = 1.53 \times 10^{51} \, \mathrm{erg}\) at the source. This represents the total jet energy required to be produced in our simulations, assuming a radiative efficiency of $\sim$10\%.

Below, we present a suite of simulation models tested against observational constraints to identify those most suitable for GRB 090510. The best-fitting models are selected based on their closest agreement with the observed $E_{\rm{GRB}}$ and $\theta_{j}$.

\subsection{Engine Parameters}

The models in our study are categorised into three groups based on disk mass: low mass (LD, $\sim 10^{-3} M_\odot$), medium mass (MD, $\sim 10^{-2} M_\odot$), and high mass (HD, $\sim 10^{-1} M_\odot$). These mass ranges align with those expected from NS mergers associated with SGRBs, providing a comprehensive basis for analysing the effects of varying disk masses. Details of models incorporating DE will be presented separately. The HD category also includes disk masses up to $0.60 M_\odot$, covering the range expected from NS-BH mergers. The magnetic field strength,  controlled by the plasma $\beta$ parameter, takes a range of $1-10^{3}$ among our models with BH spin ranging across 0.60 - 0.95. The plasma $\beta$ is normalised across the torus so that the value is maximum at the $R_{max}$ of the FM torus. 

\begin{figure}[h]
\centering
\includegraphics[width=0.45\textwidth]{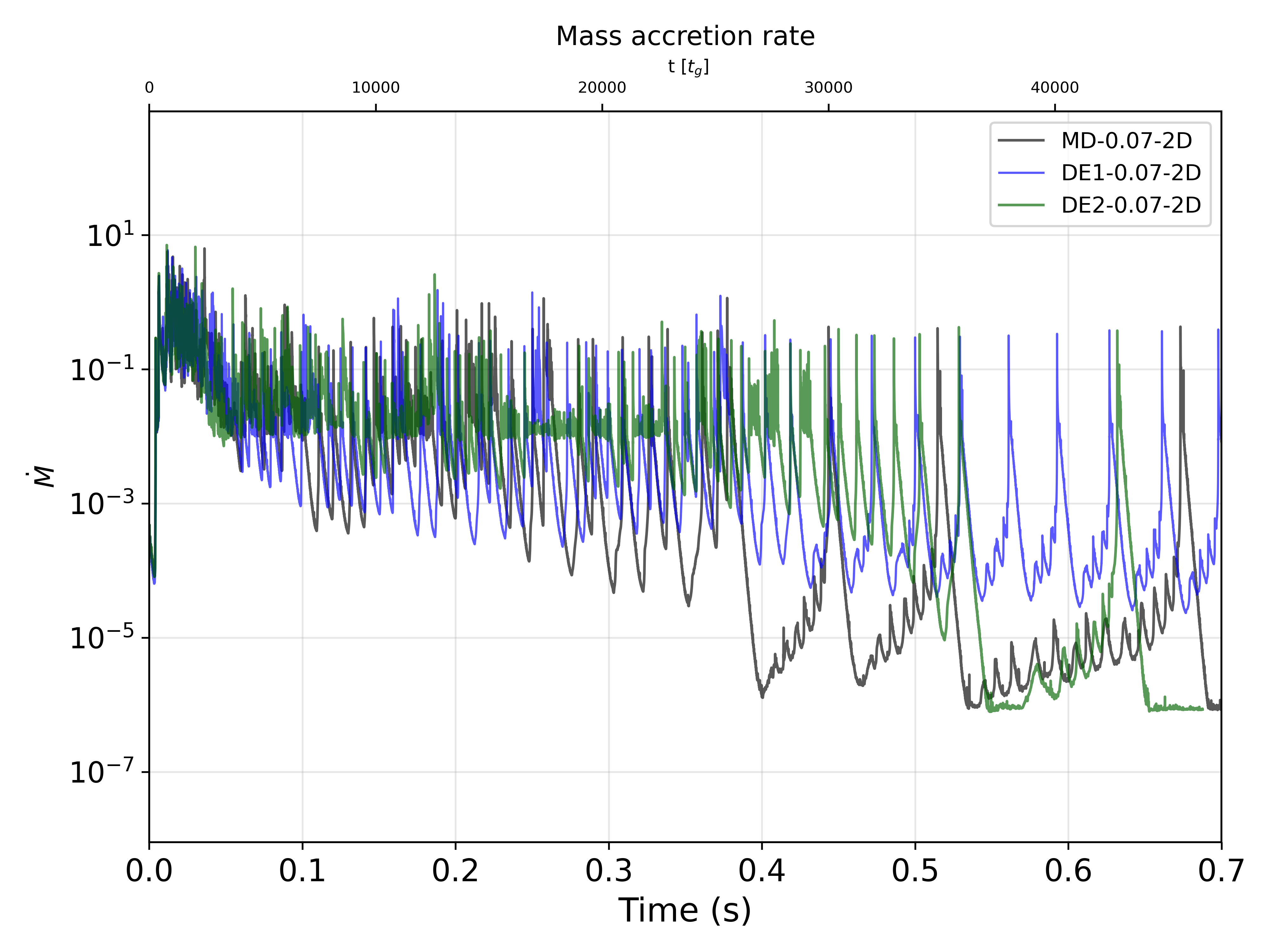}
\includegraphics[width=0.45\textwidth]{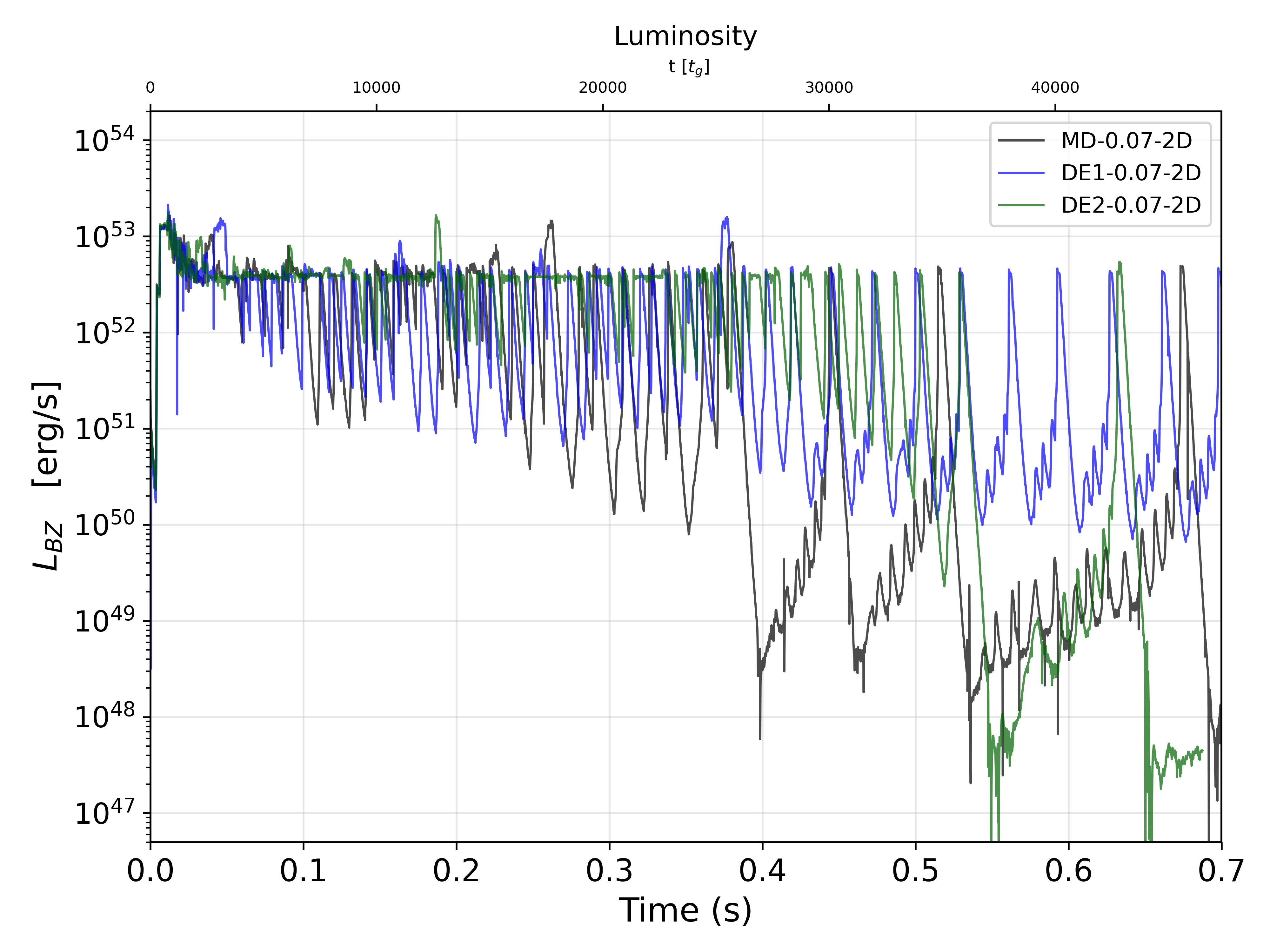}

\caption{Time evolution of mass accretion rate ($\Dot{M}$) in the upper panel and luminosity ($L_{BZ}$, lower panel) for the best models.}
\label{figure:mdot}
\end{figure}

Table \ref{tab:Sim_Models} presents comprehensive information for all models in our sample, including their initial configurations and analysed results. 
Additionally, Figure~\ref{figure:mdot} illustrates the temporal evolution of the mass accretion rate (top panel) and luminosity (bottom panel) for selected models from our simulation suite.

\begin{table*}[h]
  \centering
  \resizebox{0.95\textwidth}{!}{%
  \begin{tabular}{ccccccccccc}
    \hline
    Model & $M_{disk,I}$ &  $M_{disk,F}$ & $R_{in}$, $R_{max}$  &$a$ & $R_{out}$ & $\beta_{max}$ & $E_{jet}$  & $\theta_{jet}$ & Resolution & $t_f$ \\ 

     &  $~(10^{-3} M_\odot)$ & $~(10^{-3} M_\odot)$ & ($R_g$) & & ($R_g$) & & $(\mathrm{erg})$&(deg)   & ($N_r \times N_{\theta}\times N_{\phi}$) & ($t_g$) \\
    \hline\\
    
    \texttt{LD1-0.003-2D} & 3.130 & 2.010 &  10,12   & 0.60 & 1000 & 900  & 1.87$\times$10$^{49}$ & 25.5   & 512$\times$256$\times$1& 100k \\ 
    
    \texttt{LD2-0.003-2D} & 3.518 & 3.028 &  10,12 & 0.90 & 1000 & 1200 & 6.52$\times$10$^{49}$ & 25.6 &  512$\times$256$\times$1 & 100k \\
    
    \texttt{HD1-0.60-2D} &  623.20   & 551.10 &  50,60  & 0.60 & 1000 & 100  & 1.38$\times$10$^{52}$ & 13.5 &  512$\times$256$\times$1 & 50k \\
    
    \texttt{HD2-0.60-2D} &  626.66   & 40.59 &  50,60  & 0.90 & 1000 &3.50 & 1.70$\times$10$^{53}$ & 14.9 &  512$\times$256$\times$1 & 50k\\

    \texttt{MD1-0.06-2D} &  63.28   & 61.60 & 20,25  & 0.60 & 1000 &1600  & 3.50$\times$10$^{50}$ & 19.1 &  512$\times$256$\times$1 & 100k\\
    
    \texttt{MD2-0.06-2D} &  64.62   & 63.66 & 20,25  & 0.90 & 1000 &1600  & 1.23$\times$10$^{51}$ & 20.7 &  512$\times$256$\times$1 & 100k \\
    
    \texttt{HD-0.10-3D} & 105.67 & 39.60 & 12,18 & 0.95 & 1000 &200 & 1.10$\times$10$^{53}$ & 18.1  &  256$\times$128$\times$64 & 45k \\

    \texttt{MDC-0.08-2D} & 80.80 & 50.61 & 6,11.5  &  0.85  & 3500 & 2500 & 6.67$\times$10$^{51}$ & 8.1  &  700$\times$512$\times$1 & 40k \\
    \\
        
    \hline\\
    
    \texttt{MD-0.07-2D} & 74.32 & 41.62 & 12.5,18 &   0.80  & 3500 & 150 & 1.11$\times$10$^{52}$ & 9.2  &  700$\times$512$\times$1 & 50k\\

    \texttt{DE1-0.07-2D} & 80.97 & 45.85 & 12.5,18 &  0.80  & 3500 & 150 & 1.21$\times$10$^{52}$ &  11.1  &  700$\times$512$\times$1 & 50k\\
    
    \texttt{DE2-0.07-2D} & 86.65 & 39.87 & 12.5,18 &  0.80   & 3500 & 150 & 1.60$\times$10$^{52}$ & 10.0  &  700$\times$512$\times$1 & 50k\\\\

    \hline\\
  \end{tabular}
  }
  \caption{Models used in the simulation of GRB 090510. The $\theta_{j}$ is calculated as an average between 0.1 - 0.6~s at $R_{out}$. For $DE$ Models, total disk mass also includes contribution from the dynamical ejecta. 
   }
  \label{tab:Sim_Models}
\end{table*}

\subsection{Jet Evolution}

\begin{figure*}[ht]
    \centering
    \subfloat{\includegraphics[width=0.32\textwidth]{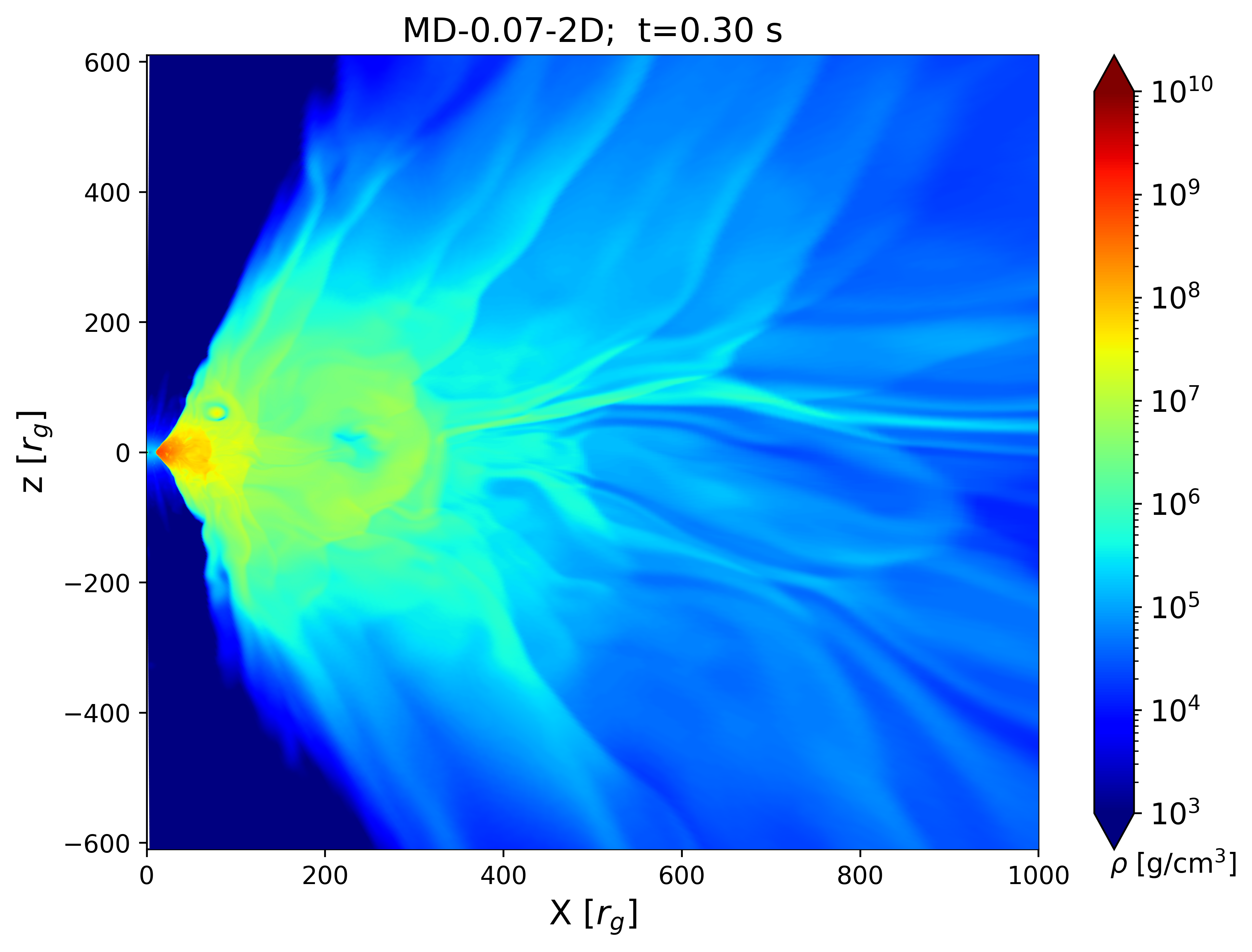}}
    \hfill
    \subfloat{\includegraphics[width=0.32\textwidth]{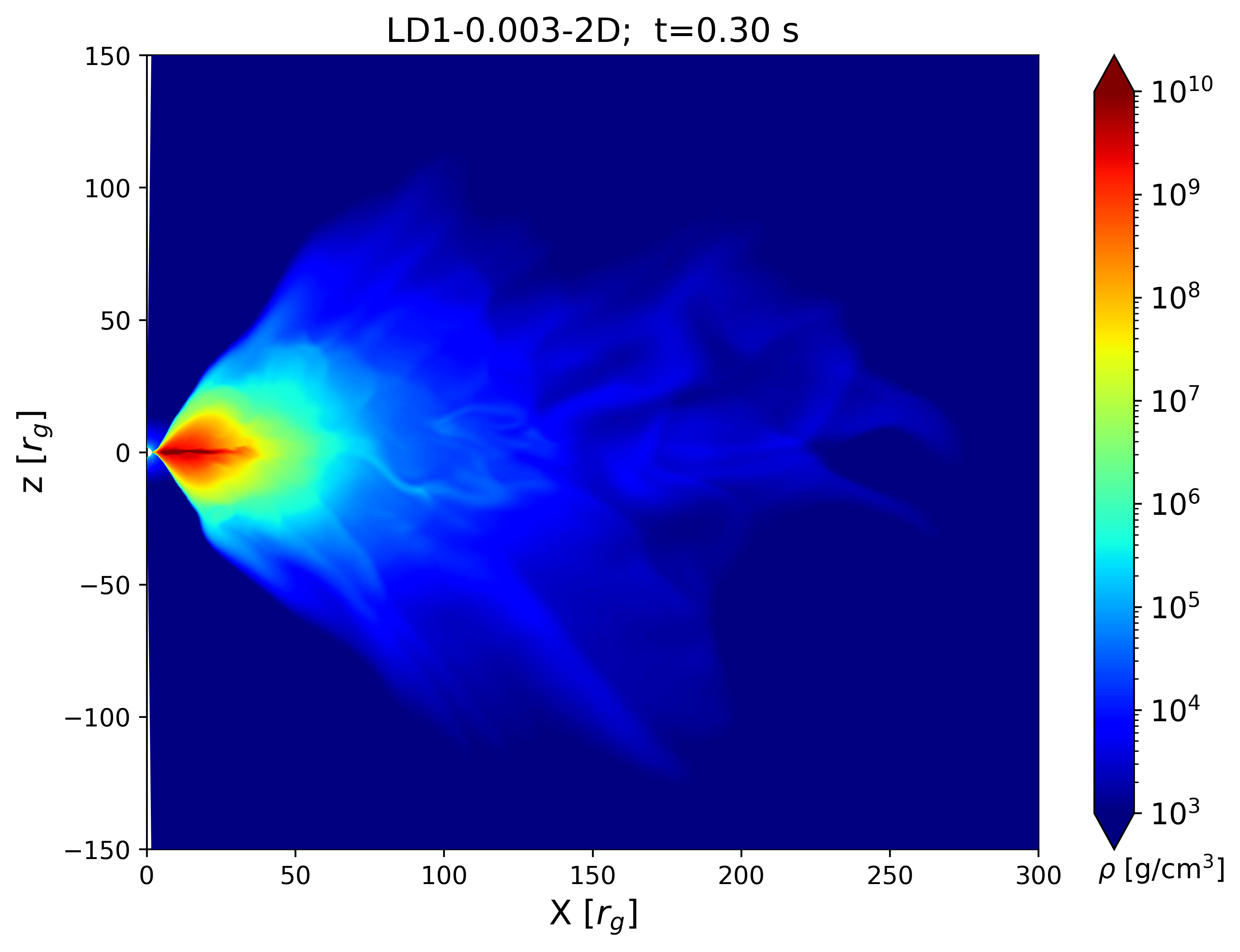}}
    \hfill
    \subfloat{\includegraphics[width=0.32\textwidth]{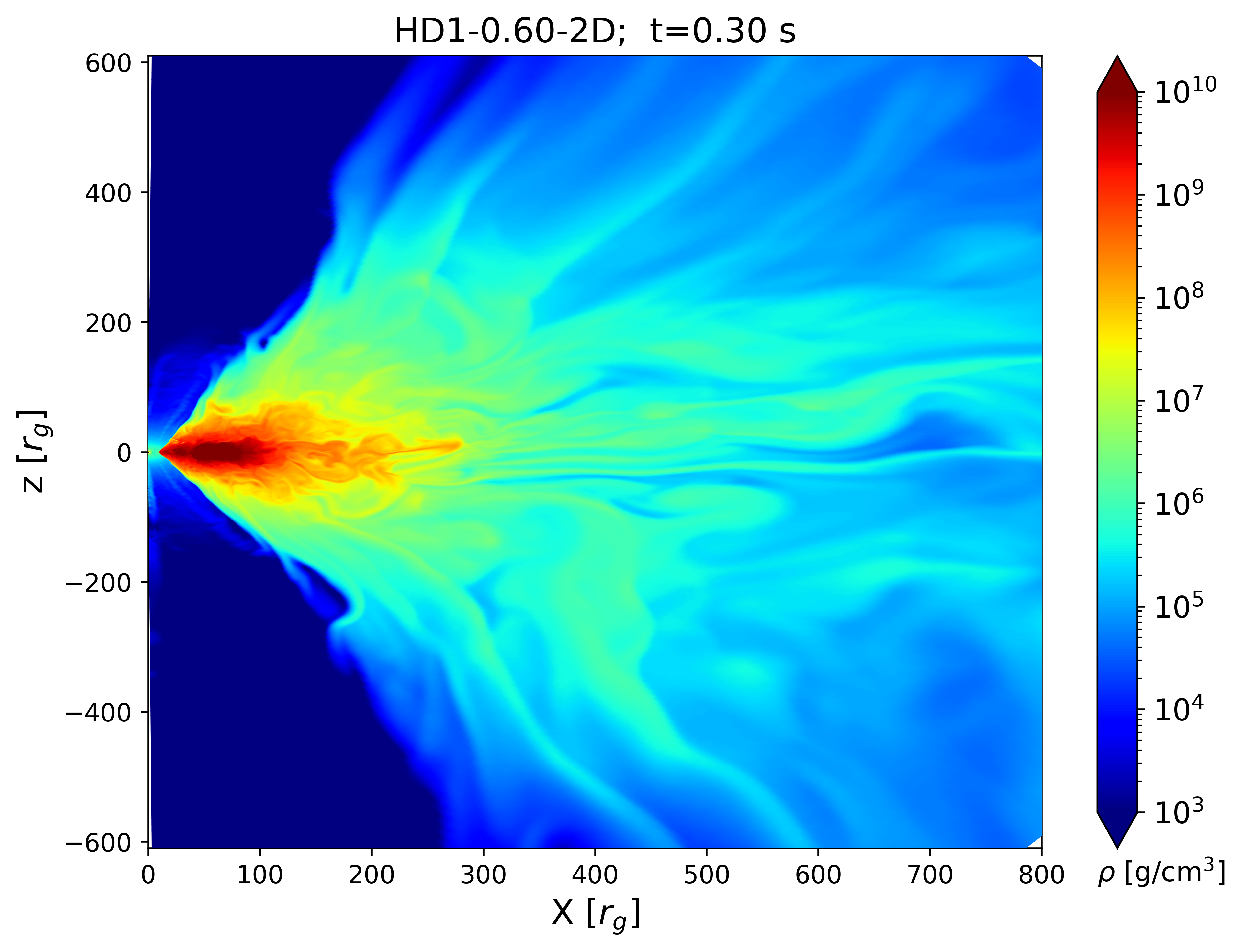}}
    \\
    \subfloat{\includegraphics[width=0.32\textwidth]{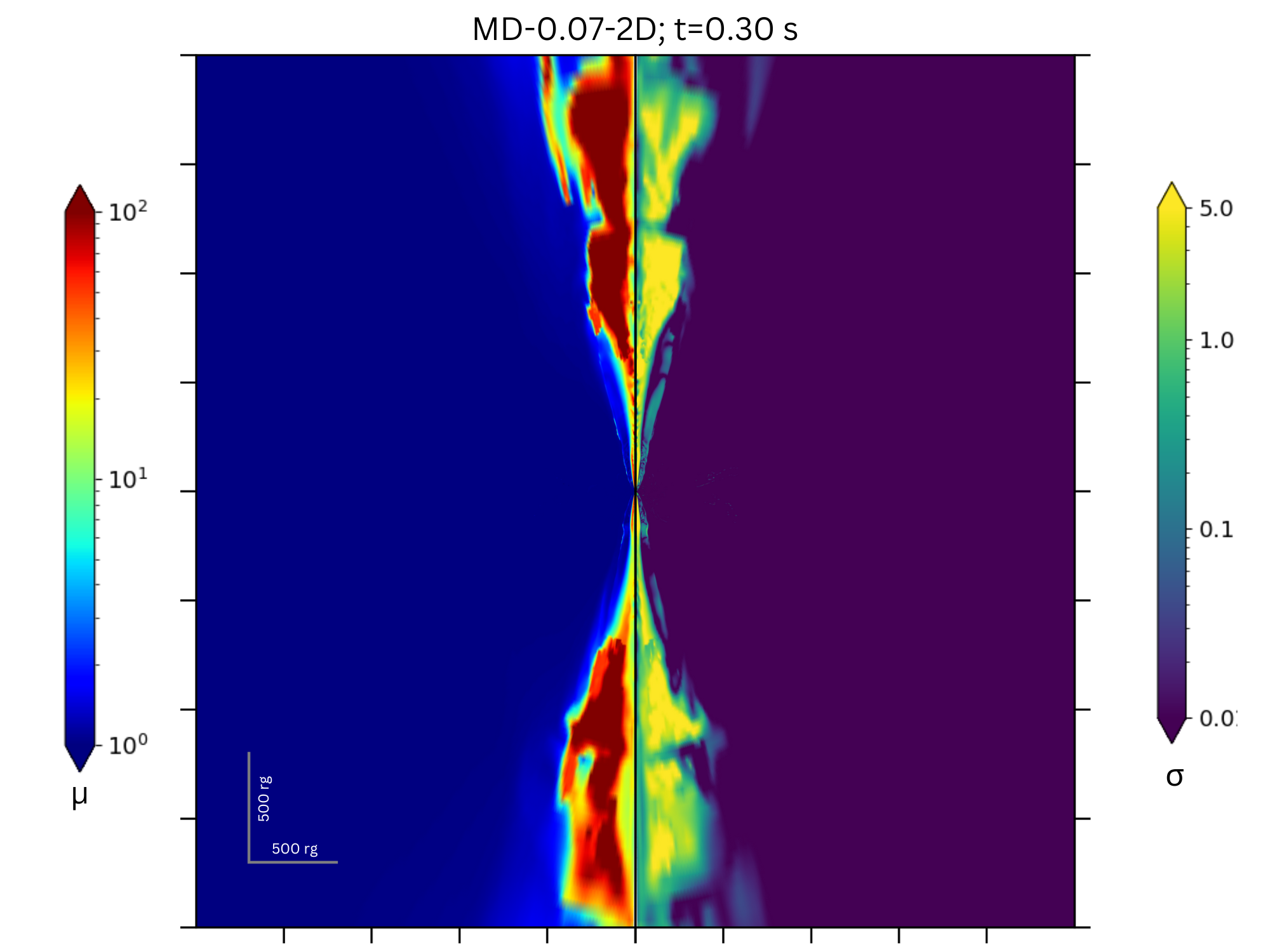}}
    \hfill
    \subfloat{\includegraphics[width=0.32\textwidth]{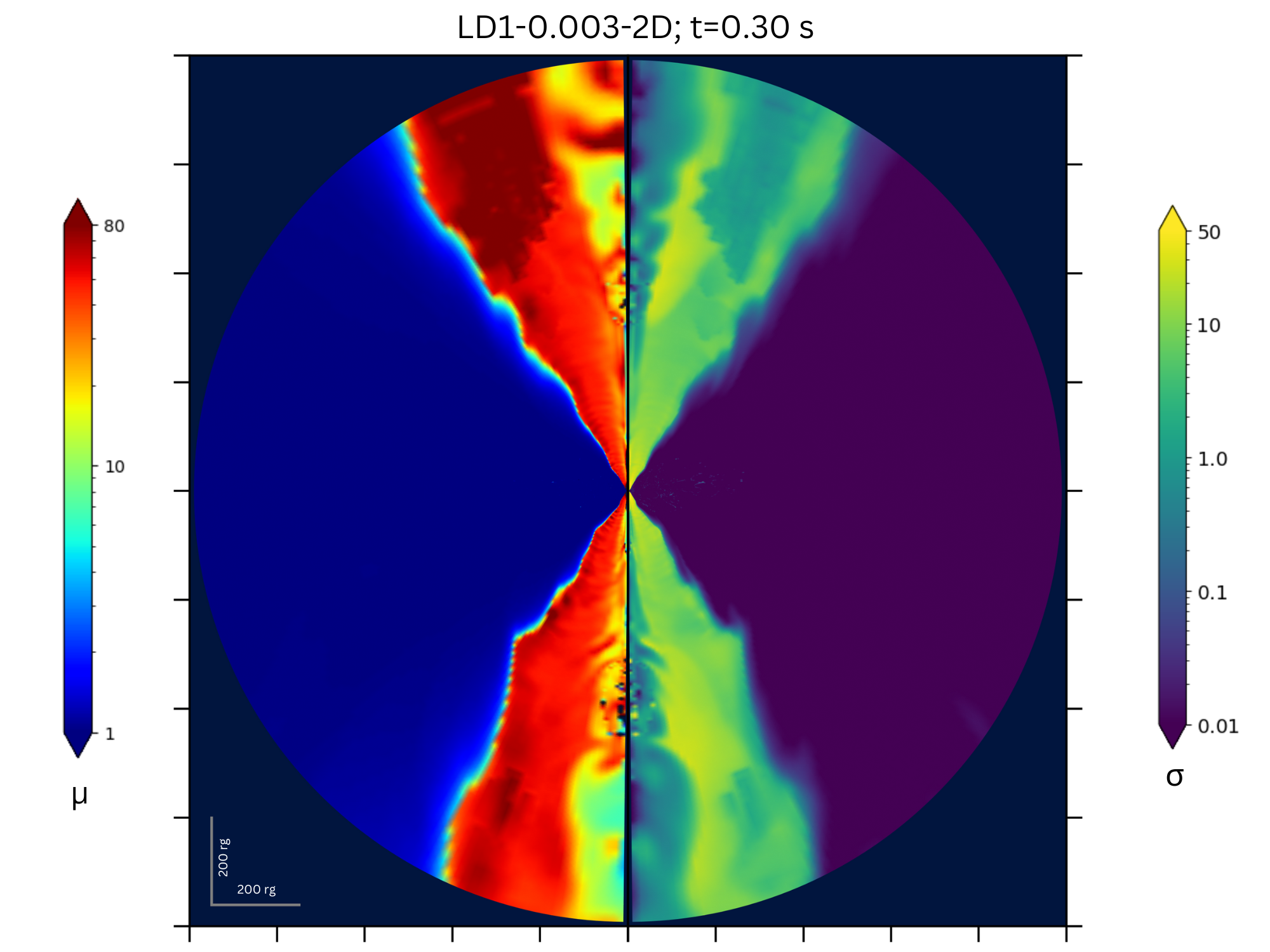}}
    \hfill
    \subfloat{\includegraphics[width=0.32\textwidth]{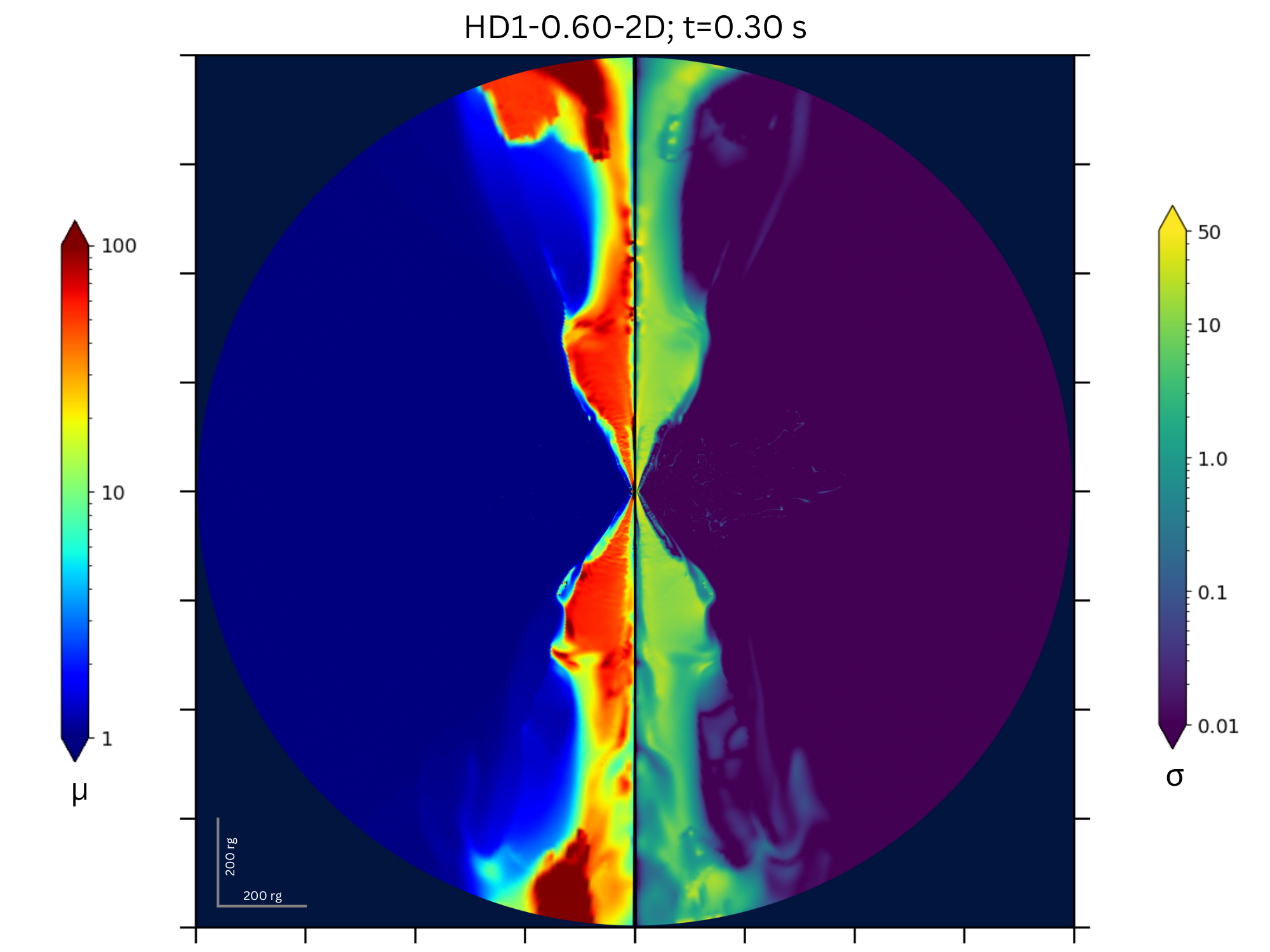}}
    \caption{Density distribution and jet structure at \(t = 0.3\) seconds for three different models. Top row: Density plots of the accretion disk illustrating variations in the disk's evolution across the three models. Bottom row:  \(\mu\) and \(\sigma\) parameters for each model, reflecting the jet energy distribution and magnetic field strength variations, respectively.}  
    \label{fig:Rho-mu-sigma-multi}
\end{figure*}

We analyse the evolution of the simulated jets, focusing primarily on the best-fit models that closely match the observations of GRB 090510. Additionally, we discuss results from other models within our suite to provide a broader perspective on short GRB jet dynamics. As the accretion process onto the BH commences, the jet is launched almost instantaneously, driven by the rapid accumulation of magnetic flux and energy extraction mechanism at BH horizon. Figure~\ref{fig:Rho-mu-sigma-multi} illustrates the density distribution of the accretion disk (top panels) and the jet structure (bottom panels) for selected models. The lower panels further depict the jet energy distribution using the $\mu$ parameter and the magnetisation profiles represented by $\sigma$.

\begin{figure*}[ht]
    \centering
    \subfloat{\includegraphics[width=0.32\textwidth]{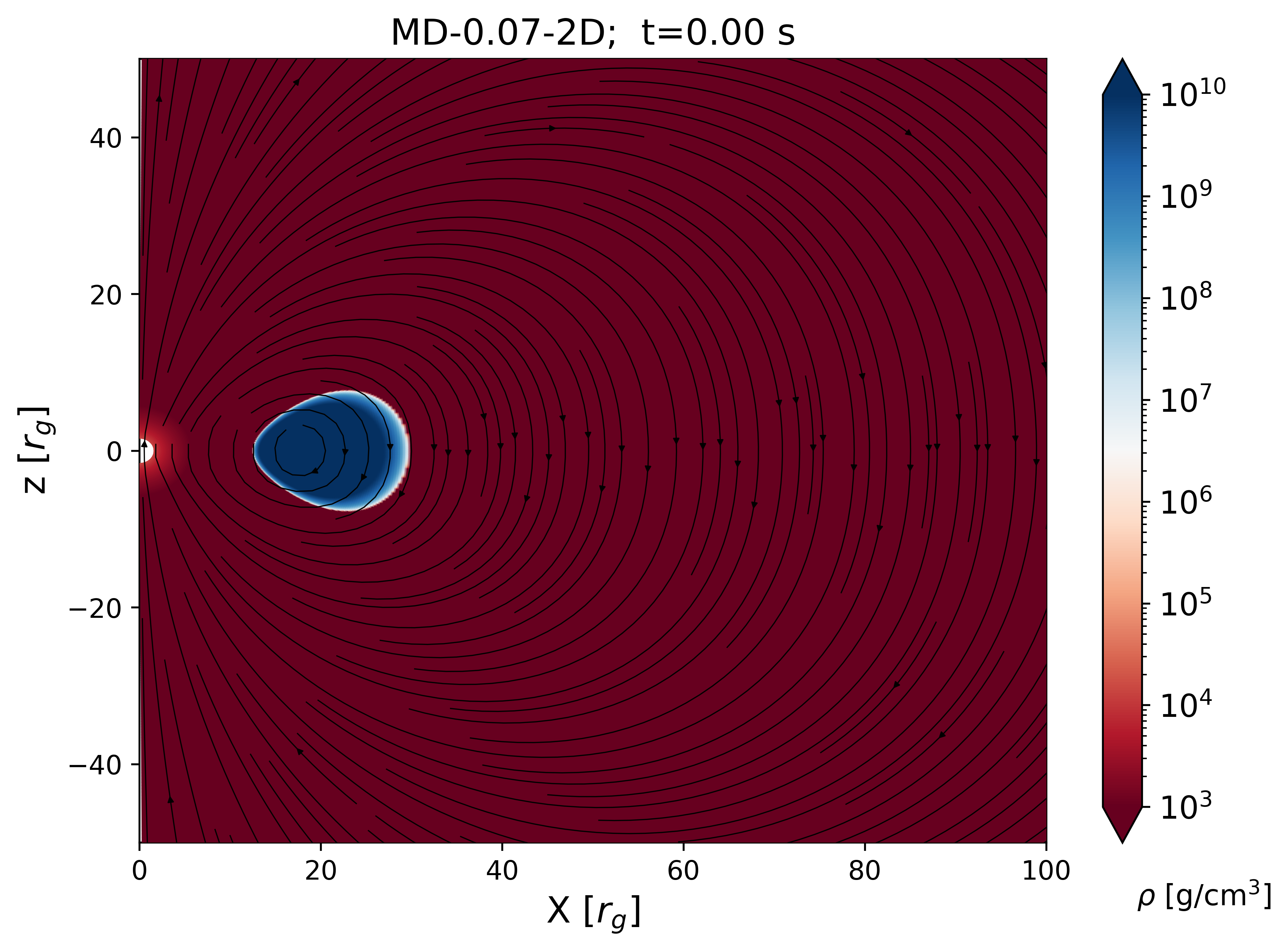}}
    \hfill
    \subfloat{\includegraphics[width=0.32\textwidth]{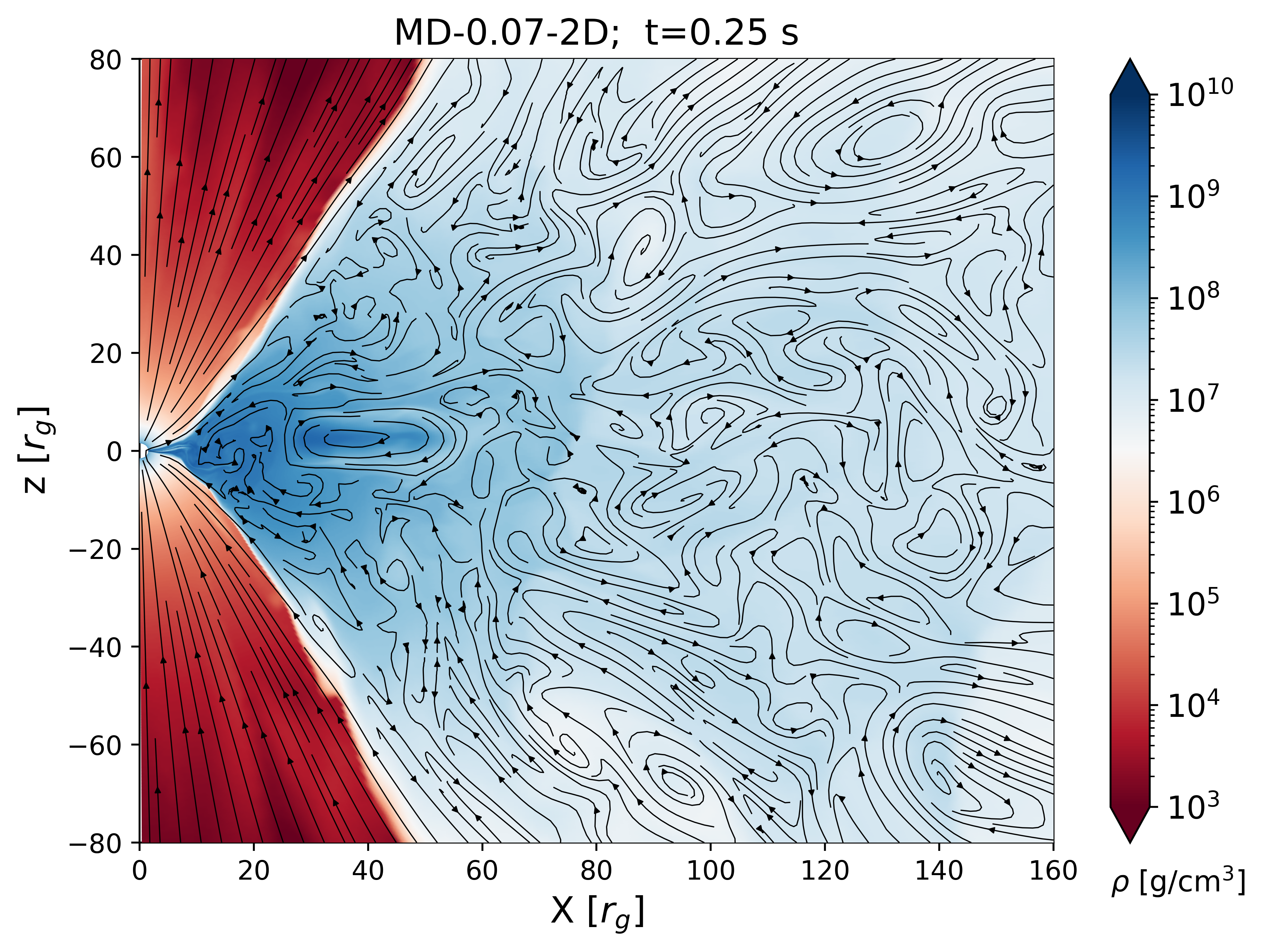}}
    \hfill
    \subfloat{\includegraphics[width=0.32\textwidth]{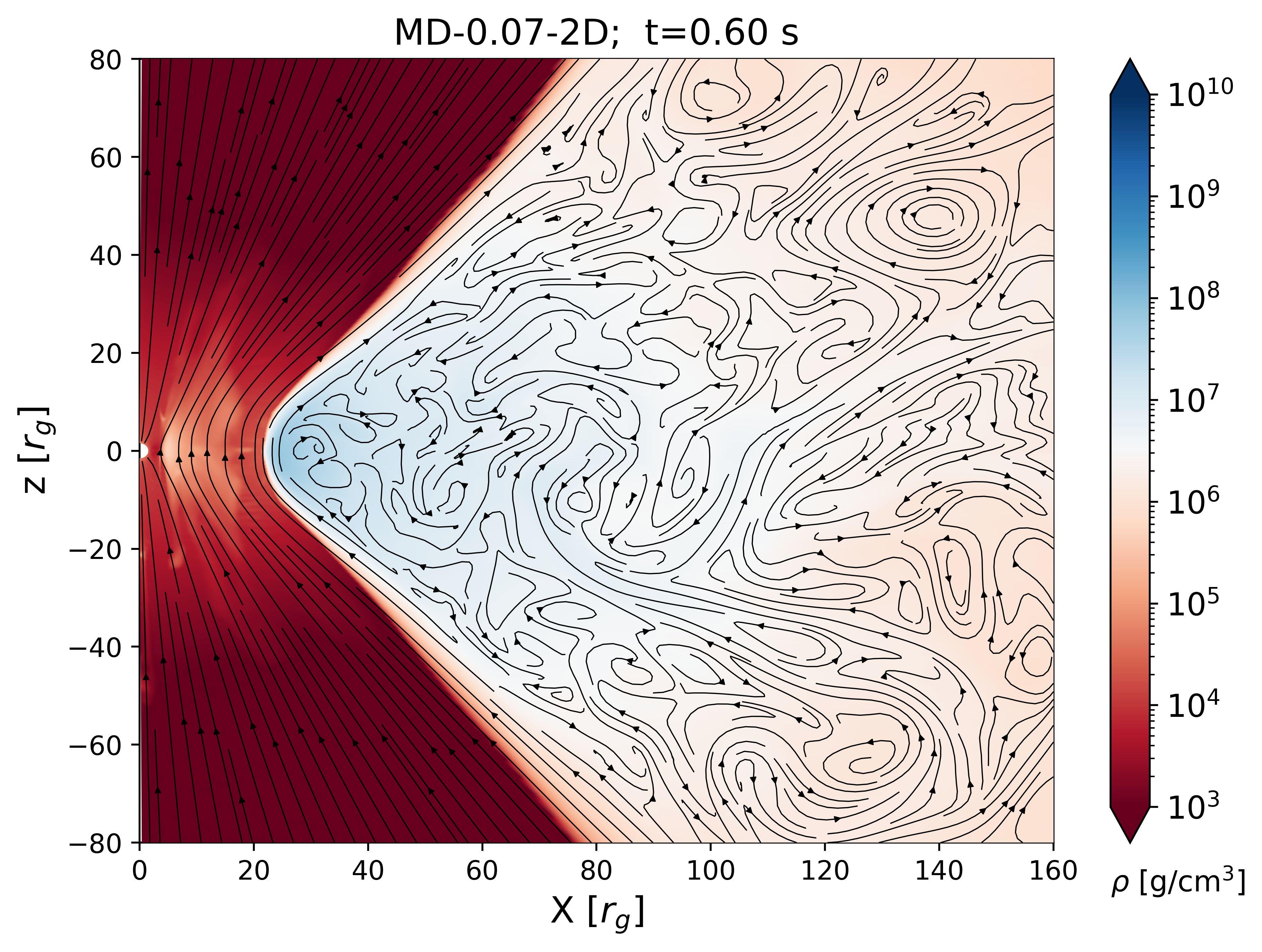}}
    \\
    \caption{ Snapshots of disk density profile  with magnetic field streamlines of model \texttt{MD-0.07-2D} at 0s (left panel), 0.25s (middle panel) and 0.60s (right panel) respectively.}
    \label{fig:Rho-B}
\end{figure*}

Among all our models, the selected best-fit models that match the energetics and $\theta_{j}$ of GRB 090510 are \texttt{MD-0.07-2D}, \texttt{DE1-0.07-2D}, and \texttt{DE2-0.07-2D}. These models share similar parameters, except that the DE models incorporate an additional radially expanding DE component, as described in Sec. \ref{Dynamic_ejecta_1}. 
All three models exhibit comparable jet properties, with average opening angles of \(\langle \theta_{j} \rangle \approx 9^\circ - 11^\circ\). The jet energy  for these models ranges between \(E_{\rm jet}\)=\( (1.1 - 1.60) \times 10^{52} \, \mathrm{erg} \). 
The density profile and jet structure for the model \texttt{MD-0.07-2D} at 0.3s is shown in the left panel of Fig. \ref{fig:Rho-mu-sigma-multi}. The evolution of magnetic fields at three distinct times (0~s, 0.25~s, and 0.60~s) for this model is depicted through streamline plots in Figure \ref{fig:Rho-B}, which effectively captures the dynamic changes in field topology. At t=0, the left panel shows the initial poloidal field configuration. The middle panel represents the field setup at t=0.25, during a stable accretion and jet phase, illustrating a well-defined magnetic structure. The right panel captures a later stage at t=0.60, where the accretion process is partially obstructed, highlighting the complex threading of field lines in the vicinity of the BH. 

Models \texttt{DE1-0.07-2D} and \texttt{DE2-0.07-2D} also have very similar evolution in both density and jet profiles. 
These models, were initialised with ejecta masses of 0.006 M$_\odot$ and 0.012 M$_\odot$, respectively, expanding radially with a constant velocity of 0.15$c$.  We adopt ejecta masses and velocities within the lower limits reported in previous numerical studies \citep{Radice_2018, 2022ApJ...933L...2G}.  All models assumed unmagnetised ejecta and a disk mass of 0.07 M$_\odot$. In both cases, the resulting jet evolution showed no significant difference in dynamics compared to the model \texttt{MD-0.07-2D}, which lacks ejecta. The estimated opening angle $\theta_j$ remained nearly unchanged. This minimal impact is likely due to the fact that a substantial fraction of the ejecta exits the computational domain shortly after simulation begins. Considering $\eta$ of 10\% for conversion of jet energy into radiation, these models can produce an average GRB energy (\(E_{\rm GRB}\)) of the order of \(\sim 1.3 \times 10^{51} \, \mathrm{erg}\), which is in close agreement with the collimation-corrected \(E_{\rm GRB}\) of GRB 090510. If we consider the value of $\eta$ as 13.7\%, 12.6\%, and 9.5\% for the models \texttt{MD-0.07-2D}, \texttt{DE1-0.07-2D}, and \texttt{DE2-0.07-2D}, respectively, we can produce the exact $E_{\mathrm{GRB}}$ of $1.53\times10^{51}$~ergs for GRB~090510.

\begin{figure}[h]
\centering
\includegraphics[width=0.45\textwidth]{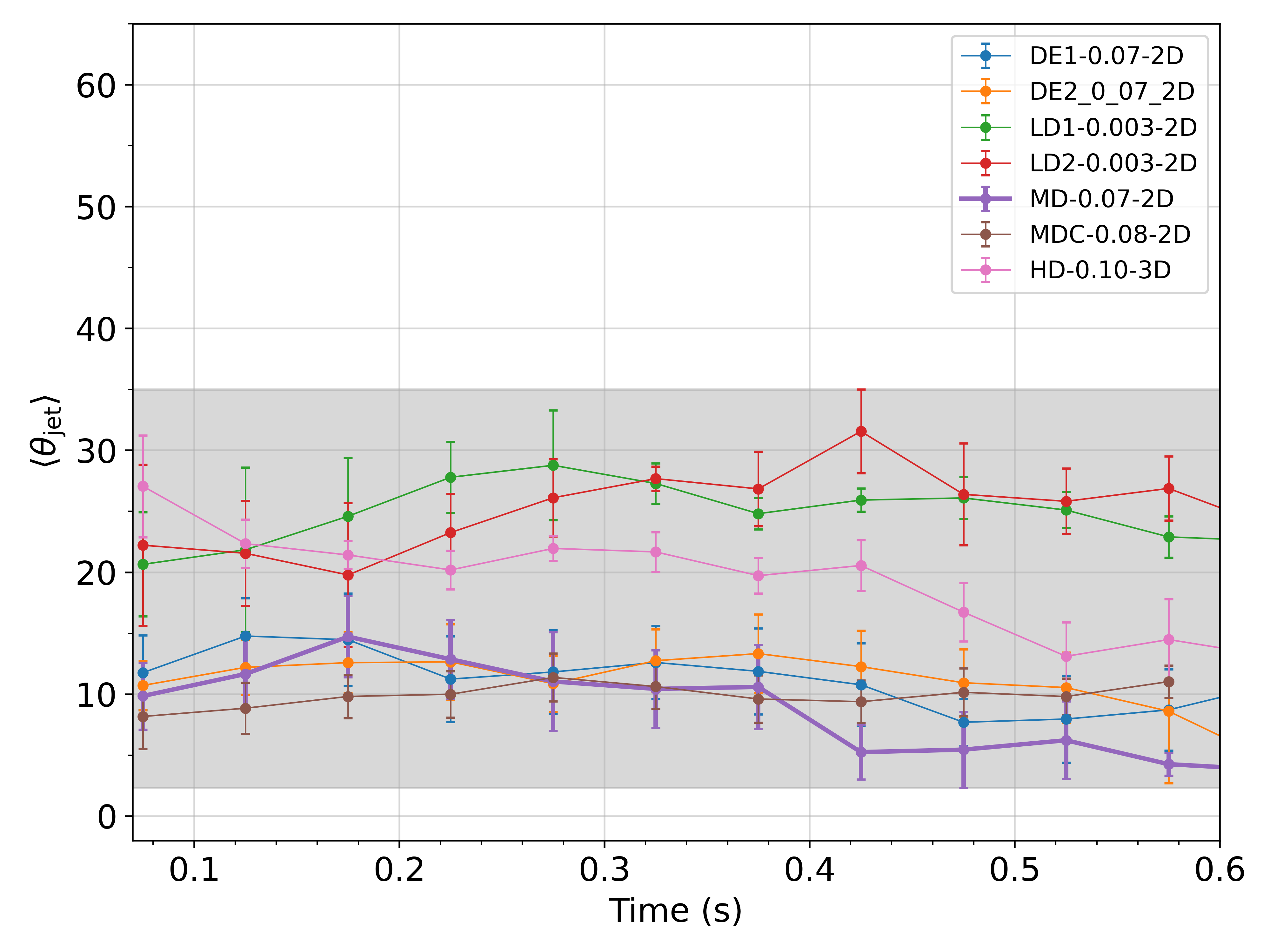}
\caption{Time evolution of the jet opening angle ($\theta_{j}$) at 1000$R_{\rm g}$ for selected models. The opening angle and their standard deviation are computed at 0.05s intervals to track its temporal variation. For clarity, only a subset of the models is shown. The grey shaded region represents the range between the highest and lowest opening angle models, each extended by its respective standard deviation.}
\label{figure:op-angle-over-time}
\end{figure}

In these models, one crucial factor that shapes the jets is the accretion disk winds - outflows of matter propelled outward by magnetic and thermal forces from the disk, significantly affecting the dynamics and emissions as the jet propagates away from the central engine. The temporal evolution of $\theta_{j}$ for these models is provided in Fig. \ref{figure:op-angle-over-time}. This shows, on average, how jet structure changes over time. 

Models \texttt{LD1-0.003-2D} and \texttt{LD2-0.003-2D} represent the low-density disk configurations in our sample. Although these models do not capture the specific observed properties of GRB 090510, they provide valuable insights into SGRB jet behaviour under the assumption of a less dense accretion disk. The impact of disk winds on the structural changes in the jet for these models is minimal, which is evident from the density and jet structure shown in the middle panel of Fig. \ref{fig:Rho-mu-sigma-multi}. This minimal impact of disk winds results in the largest $\theta_{j}$ ($\sim25^0$) among all configurations in our study (see Table \ref{tab:Sim_Models}). 
Furthermore, the simulations for these models were run for over two seconds, demonstrating stability and producing a consistent luminosity of the order of $10^{49} \, \mathrm{erg \, s^{-1}}$ until the end of the simulations.

The models \texttt{HD1-0.60-2D} \& \texttt{HD2-0.60-2D} have the highest disk mass among our sample, accompanied by a highly turbulent and discontinuous jet structure throughout their lifetimes. See the right panel of Fig. \ref{fig:Rho-mu-sigma-multi} for the density and jet structure for the model  \texttt{HD1-0.60-2D} at 0.3 seconds. 
Model \texttt{HD2-0.60-2D} also has the lowest plasma $\beta$ among all our samples, producing a jet with a total energy of $E_{\rm jet} = 1.70 \times 10^{53} \, \mathrm{erg}$, highest among all our models.
Notably, the most stable model with the lowest $\theta_{j}$ is \texttt{MDC-0.08-2D}, which has an FM torus with (\(R_{\rm in}, R_{\rm max}\)) = (6, 11.5) \(R_g\). This model represents the case with closest distance between the initial torus  and the BH in our study. Despite having a high plasma \(\beta\) of 2500, which was implemented to observe jet evolution over longer durations, this model produces a stable jet with an opening angle of $\theta_{j}$\(\sim 8^\circ\) throughout its lifetime. This effect can be attributed to several key factors: a smaller disk radius results in a stronger interaction between the jet and disk winds, which collimates the jet at an earlier stage.
Also, a high plasma-\(\beta\) environment reduces magnetic instabilities, allowing for a more stable jet structure; and the immediate jet-wind interaction at injection time helps maintain a narrow outflow. In contrast, when we attempt to increase the magnetic field strength to achieve a jet energy comparable to our target GRB, the jet becomes highly unstable. This suggests that while proximity to the BH aids collimation, achieving a balance between magnetic field strength and jet stability remains a key challenge for reproducing realistic GRB jets. The effect of disk geometry and other properties on jet structure and dynamics, especially in collimation, is discussed in detail in previous works \citep{10.1093/mnras/stz2552, Hurtado2024ApJ...967L...4H}.

\begin{figure}[h]
\centering
\includegraphics[width=0.45\textwidth]{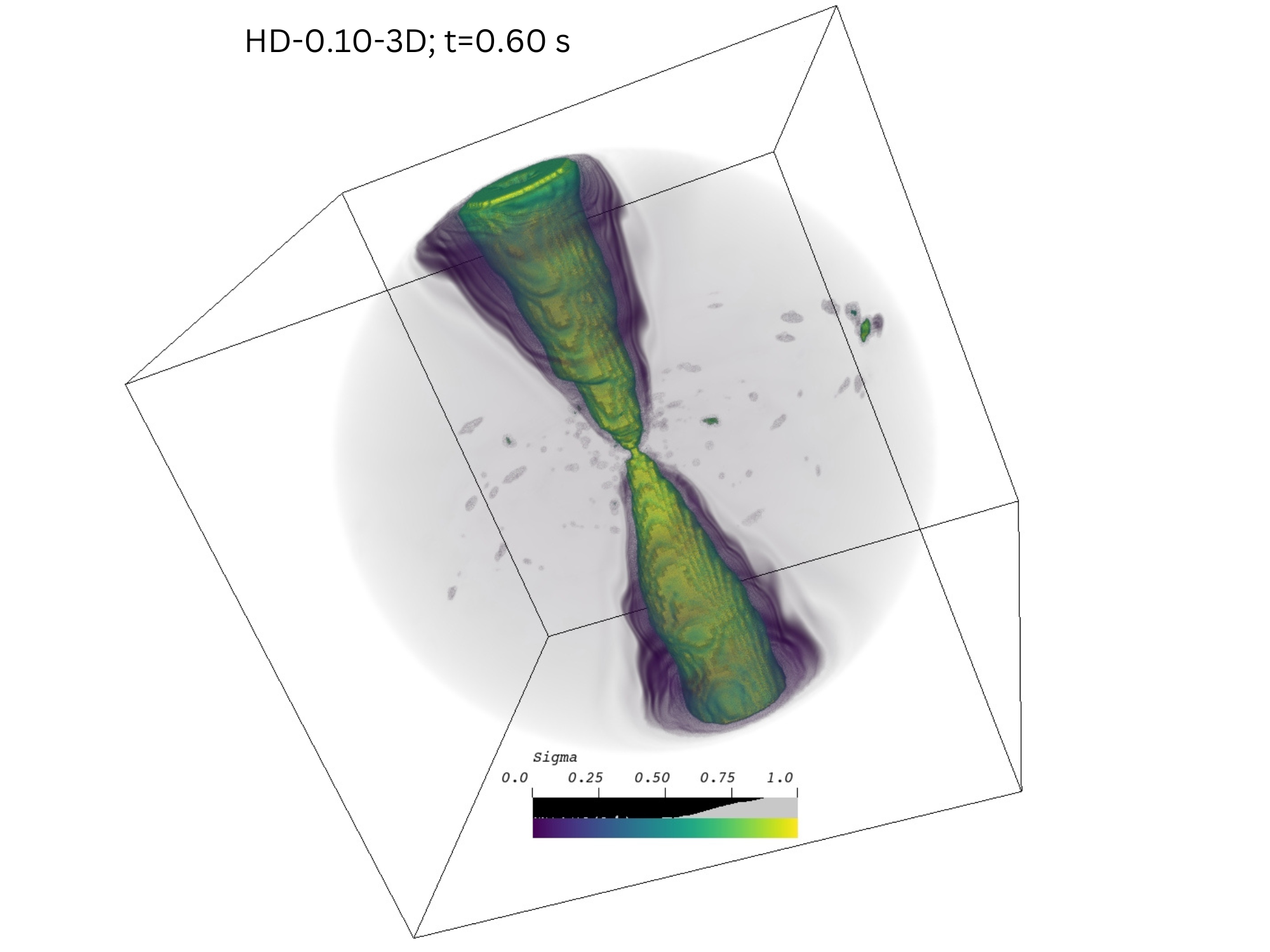}
\caption{Jet structure of 3D model in our sample, HD-0.10-3D. The parameter represented is jet magnetisation $\sigma$, and the bounding box is 500~$R_g$.  }
\label{figure:3D-Sigma}
\end{figure}

In our sample, we have included one 3D model:  \texttt{HD-0.10-3D}. 
This model was evolved until 0.6 seconds, yielding an average jet opening angle of $\theta_{j} \sim 18^{\degree}$. The opening angle is estimated and averaged over each azimuthal ($\phi$) slice. Although this value is higher than that observed for GRB 090510, the jet gradually collimates over time, reaching approximately $\theta_{j} \sim 13^{\degree}$ toward the end of the simulation. This collimated state of the jet from this model is shown by the 3D volume rendering map of jet magnetisation, $\sigma$, in Fig. \ref{figure:3D-Sigma}. 
From the results, we expect the jet to be more collimated with time. The three 2D models that we found to be most compatible with the observed target GRB also show a gradual increase in their collimation over time. On the other hand, \texttt{HD} models with strong magnetic fields and wind disturbances have no specific pattern and less predictable opening angle evolution.

In total, we analysed 11 models incorporating diverse initial simulation configurations, yielding jets with a broad range of observable properties. Table \ref{tab:Sim_Models} provides an overview of all models, including their initial setups and average observable parameters. Our discussion here concentrates on selected models from each category, particularly those capable of reproducing the properties of GRB 090510.

\subsection{Lorentz Factor Evolution}
\begin{figure}[h]
\centering
\includegraphics[width=0.45\textwidth]{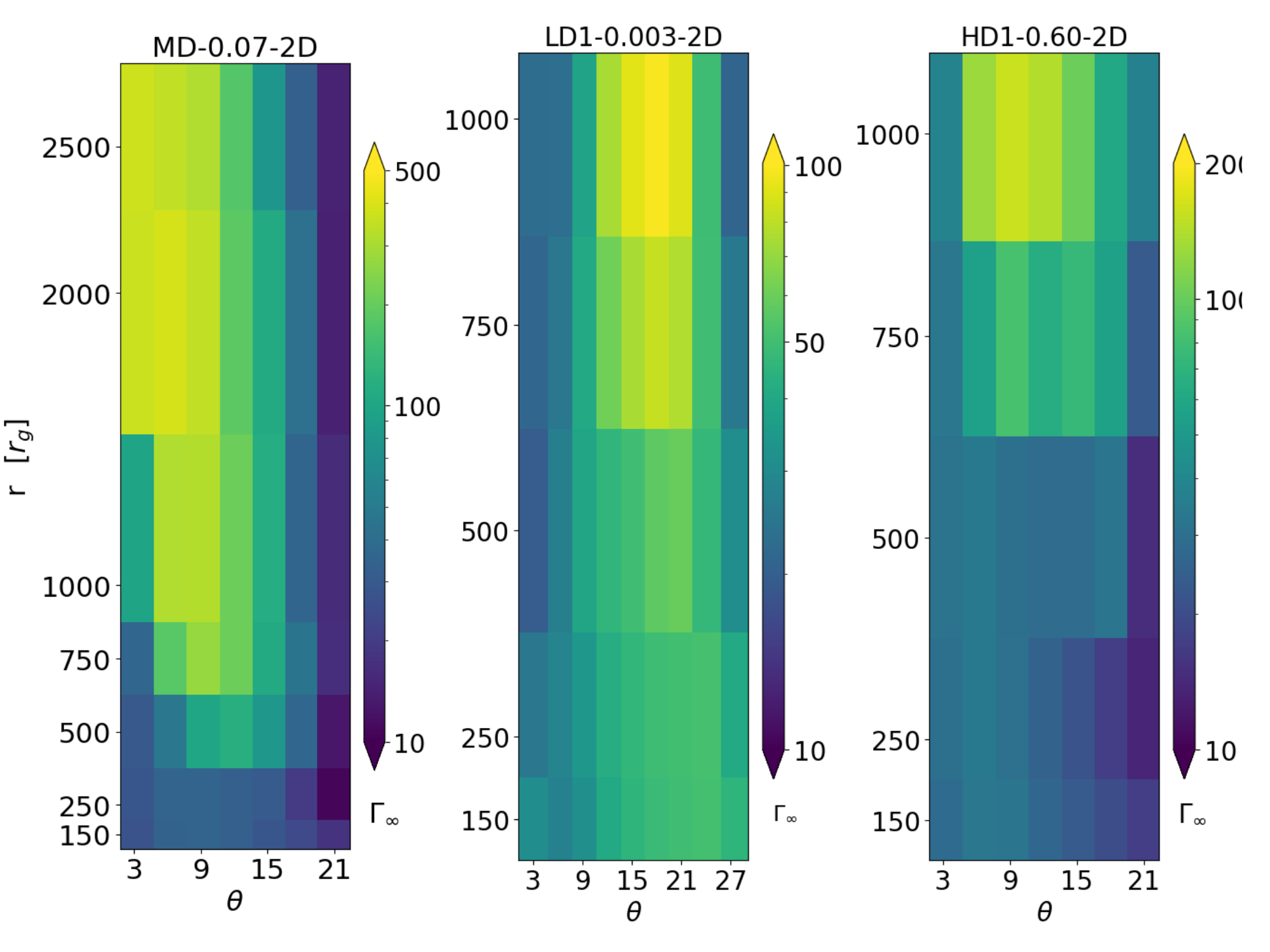}
    \caption{Visualization of the $\Gamma_{\infty}$ distributions across different radii and angular ranges. The figure comprises three panels, each representing the $\Gamma$ factor as a function of radius (r) on the y-axis and angle (\(\theta\)) on the x-axis for three different models.}
\label{figure:Loremtz-Factor-Mesh}
\end{figure}

\begin{figure}[h]
\centering
\includegraphics[width=0.45\textwidth]{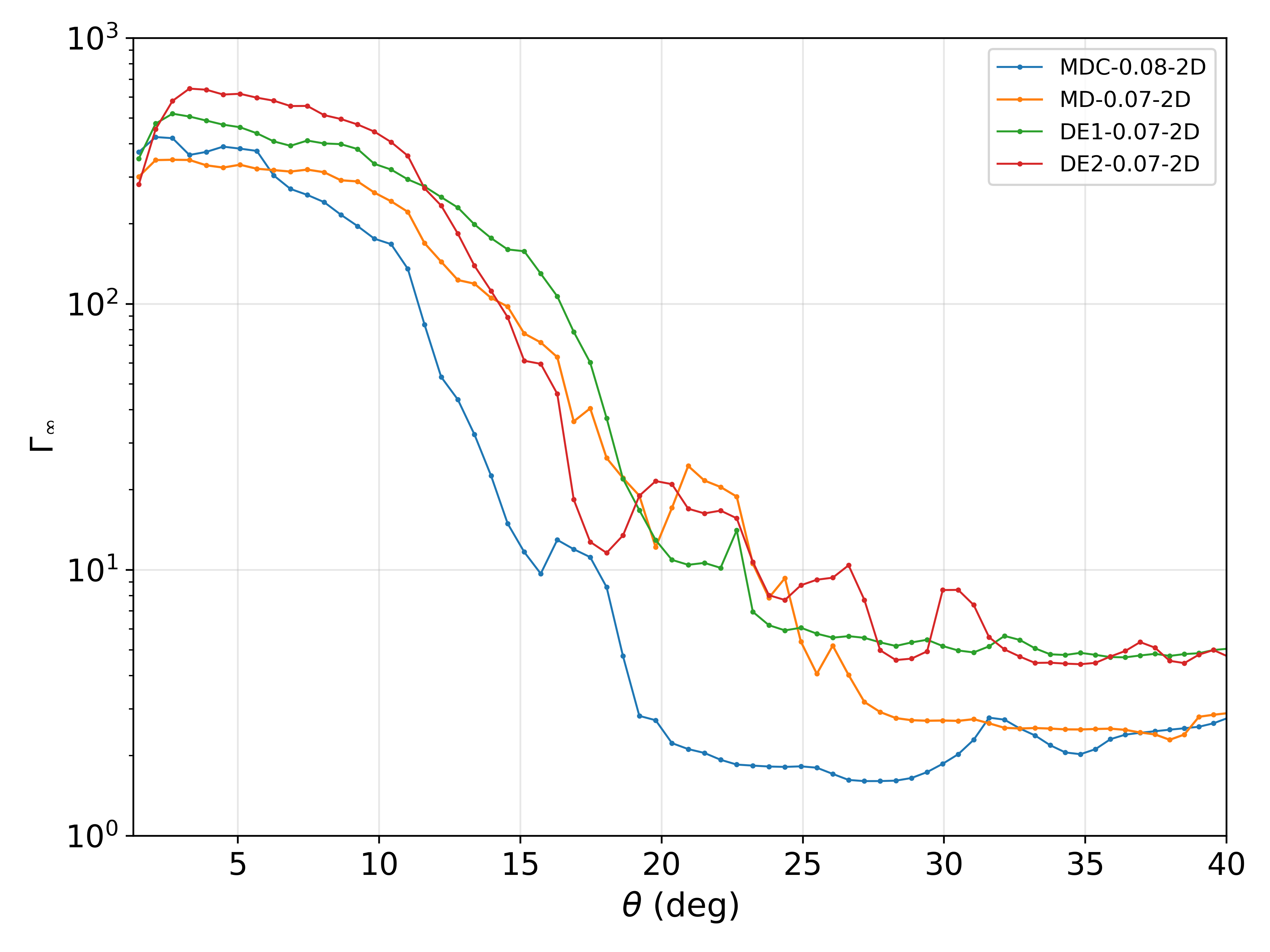}
\caption{Variation of Lorentz factor ($\Gamma_{\infty}$) as a function of polar angle $\theta$ for best-fit models, measured at a radius of 3000 $R_g$}
\label{figure:Gam_vs_th}
\end{figure}

As discussed in Sec. \ref{sec:jet_characteristics}, we calculate the terminal Lorentz factor $\Gamma_{\infty}$, 
by taking the time average of $\mu$. For selected models, we have presented the $\Gamma_{\infty}$ factor at different regions of the jet in Fig. \ref{figure:Loremtz-Factor-Mesh}. This visualisation highlights the spatial variability of the Lorentz factor across the jets, underscoring the distinct behaviours and characteristics of each model's jet structure.
We could also observe the existence of a hollow core in the jets for an average angle of $3^\degree-5\degree$ which later gets closed over larger distances. Notably, our model \texttt{MD-0.07-2D}, which aligns closely with GRB 090510, exhibits $\Gamma_{\infty}$ factor estimates reaching as high as 375 at 3000$R_g$ and continues to increase with radius. The $\Gamma_{\infty}$ as a function of polar angle $\theta$ at a fixed radius of 3000 $R_g$ for the \texttt{MD} models in the samples is provided in Fig. \ref{figure:Gam_vs_th}.

\subsubsection{Jet Variability From Simulations}

In our simulations, the jet energetic parameter $\mu$ provides a measure of the total energy content of the jet. The time evolution of $\mu$ can give insights into the dynamics of processes within the jet that lead to variability, which we observe in the GRB LCs. Various studies have commented previously regarding jet variability from simulations \citep{10.1111/j.1365-2966.2008.13425.x, 10.1093/mnras/stw1366, Janiuk_2021, James_2022, Pais2024ApJ...976...35P}. 
We note that detailed calculations of the photospheric emission are needed to address the observed variability and non-thermal spectrum that arise in observations. However, even within the current setup, we are able to find hints of this variability in parameters like $\mu$ and $\sigma$.

\begin{figure}[h]
\centering
\includegraphics[width=0.45\textwidth]{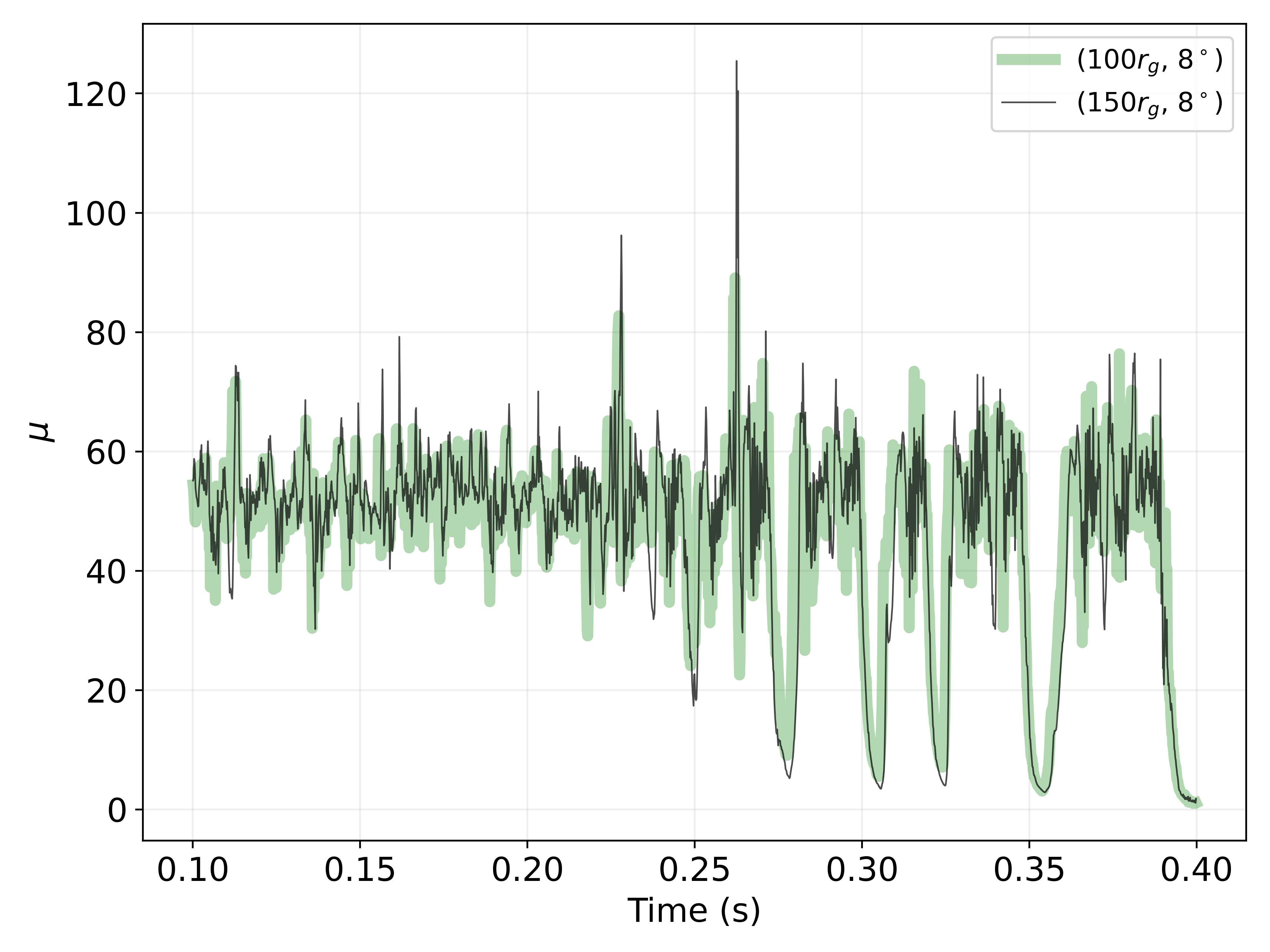}
\caption{Time evolution of the jet energetic parameter $\mu$ over time at two selected regions in the jet for the model \texttt{MD-0.07-2D} over which we have done the MTS calculations.}
\label{figure:mu_time_profile}
\end{figure}

The time evolution of $\mu$ parameter is provided in Fig. \ref{figure:mu_time_profile}  at two different chosen locations (100 $R_g$, $8^{\degree}$), (150 $R_g$, $8^{\degree}$) for the model \texttt{MD-0.07-2D}. The parameter exhibits significant variability, which we use to investigate jet's temporal evolution. Through our analysis of the  $\mu$ parameter, we can compute the theoretical MTS, which is comparable with the MTS from observations as described in Sec. \ref{MTS-obser}. By following the methodology described in \cite{Janiuk_2021} \& \cite{James_2022}, by computing the full width at half maximum (FWHM) for each discrete pulse and subsequently deriving a mean measure, we find variability timescales of 4.75~ms \& 4.98~ms respectively in two locations [(100 $R_g$, $8^{\degree}$) \& (150 $R_g$, $8^{\degree}$)] in the grid. These values are comparable to the estimated MTS for GRB 090510  from \cite{2013MNRAS.432..857M} using wavelet analysis, which is $4.9^{+1.1}_{-0.9}$ ms. The comparison between the model and observations is within 1 $\sigma$ in both locations in the grid.

\subsection{Application to additional short GRBs}

To demonstrate the broader applicability of our results beyond GRB 090510, we identify three additional short GRBs from the Fermi-GBM 10-year spectral catalog \citep{Poolakkil_2021}, for which E$_{\rm iso}$ estimates are available and jet opening angles can be inferred using the method described in section \ref{sec:jet-opening}. Redshift values for these bursts are taken from \citet{https://doi.org/10.48550/arxiv.2501.03614}. These GRBs are listed in Table \ref{tab:more-GRBs}, along with their E$_{\rm iso}$, redshift, inferred $\theta_{\rm jet}$, and the best-matching simulation models from Table \ref{tab:Sim_Models}. The simulated and observed values show good agreement (within 10\%), in both energetics and opening angle. The collimation-corrected energies for these GRBs are also consistent with our models, based on Eq. \ref{eq:ejet} and when adopting the radiation efficiency in the range of 1–15\%.

To further explore the consistency of our models with jet parameters derived using independent methods, we include GRB~120804A in our comparison. This burst has an opening angle estimated from afterglow modelling \citep{Fong_2015}. The derived $\theta_{\rm jet}$ and energetics for this GRB are in close agreement with our simulated models. This reinforces that, for observationally reasonable angle estimates, our simulation framework can successfully reproduce key jet properties. 
This extended comparison illustrates that our simulation suite spans a realistic parameter space capable of reproducing not only GRB 090510 but also a broader population of short GRBs with similar jet properties.

\begin{table}[htbp]
\centering
\begin{tabular}{lcccl}
\hline
GRB & z & \({E_{\text{iso}}}\) [erg] & \({\theta_{\text{jet}}}\) [deg] & Model \\
\hline
150101B   & 0.13 & \(4.2 \times 10^{48}\) & 25.7 & \texttt{LD1-0.003-2D} \\
090927    & 1.37  & \(1.21 \times 10^{51}\) & 21.1  & \texttt{MD2-0.06-2D} \\
100117A    & 0.92  & \(9.75 \times 10^{50}\) & 18.60 & \texttt{MD1-0.06-2D} \\
120804A$^{\ast}$   & 1.3   & \(3.4 \times 10^{52}\) & $\geq13$ & \texttt{HD1-0.60-2D} \\
\hline
\end{tabular}
\caption{Jet opening angle and energetics for additional short GRBs used for comparison with our simulation models.  The data for first three sources are from \citep{Poolakkil_2021}. The last source 120804A has opening angle estimated from multi-wavelength observations \citep{Fong_2015}.}
\label{tab:more-GRBs}
\end{table}

\section{Discussion}\label{Discussion and conclusions}

The primary goal of this work is to identify theoretical parameters that reproduce key observational features of the high-energy GRB 090510. To this end, we performed a suite of GRMHD simulations, focusing primarily on 2D runs. Additionally, we tested a single 3D configuration for completeness, though this preliminary attempt did not closely match the observed properties of GRB 090510. Given their significantly higher computational cost, the 3D simulations remain preliminary and are not the central focus of our present analysis. Instead, we base our conclusions largely on the 2D results, which provide an approximate range of jet properties for comparison with observational constraints.

However, in the literature as well, the discussion of theoretical models is hardly matched directly with observations due to the limited computational resources typically available.  In this work, we mainly focused on matching the simulated energy and opening angle of the GRB jet, which fits exactly the observations within 1 $\sigma$ (if we follow the efficiency level between 9.5-13.7\%).

An important point to consider in the GRMHD simulations is the magnetic field strength and its initial configuration. This feature can directly shape the jet profile. Some authors consider toroidal fields in the initial state of the disk-jet simulation, e.g. \cite{10.1093/mnras/stz2552}. Such fields must, however, originate from the NS merger system. The poloidal field that is adopted in our work is still in agreement with the outcome of numerical relativity simulations \citep{Paschalidis2015ApJ, Sapountzis_2019}. 

Most of the models studied in this work required a high magnetisation to produce targeted high jet luminosity. We find the existence of a Magnetically Arrested Disk (MAD) state in these models. The MAD state strongly affects the creation of the outflow and the jet dynamics. In the 2D simulations, the evolution of magnetic fields is limited by the anti-dynamo theorem.
The central engine's time variability is strong but with highly periodic accretion cycles. These cycles lead to luminosity changes and quiescent intervals, and hence, jet launching is sometimes suppressed. 
Since the non-axisymmetric instabilities \citep[e.g., Rayleigh-Taylor, kink instabilities, see;][]{Bromberg_2019}
are suppressed, the jet emission is unstable. Therefore, the overall opening angle estimates in our analysis will differ between the 2D and 3D simulations. In our simulations, the effect is prominent in HD models, while it seems to be comparably lower in LD and MD models (they are in the standard and normal evolution (SANE) mode most of the time). 
In the 3D simulations, the MAD state is easily resolved, and we could observe the interchange instabilities or magnetic reconnections \citep{Janiuk2022, Nalewajko2024}. This leads to a more efficient accretion and evolution of the accretion rate, reducing the amplitude of variability seen in 2D simulations. 

The role of dynamical ejecta in shaping jet propagation has been highlighted in several recent studies. Analytical and numerical models suggest that interaction with dense, radially expanding ejecta can lead to significant jet collimation on short timescales \citep{2020MNRAS.491.3192H,10.1093/mnras/staa3276, Nagakura_2014, Pais_2023, 10.1093/mnras/staf377}. In our simulations, however, this effect could not be confirmed due to computational limitations. The restricted radial extent of the simulation domain causes both the ejecta and the jet head to leave the grid within milliseconds of jet launching, preventing the formation of a stable collimation structure. While our setup provides a first-order test of dynamical ejecta influence, more extended simulations with larger domains and finer resolution will be required to fully capture its effect on jet evolution.

Continuing on our quest to match the observations and the theory, we have investigated the variability parameter.
The time variability has been discussed in many avenues in the literature, both for GRBs and active galactic nuclei (AGNs), and it is subjected to several definitions. Hence, there are different ways in which variability can be computed. As already mentioned, it is beyond the scope of the current paper to discuss which method is more appropriate, but it is striking to observe that the calculation of the wavelet analysis is compatible with our simulations within 1 $\sigma$. Given that the scope of the current paper is to test the reliability of this model and discuss its possible limitations, we focus on the main parameter which is more univocally reproducible, such as the isotropic luminosity of the GRB, $L_{iso}$ which is compatible within 1 $\sigma$ with our estimate. Indeed, our estimate of the observational bolometric luminosity (which is a proxy of the actual bolometric luminosity) is derived by summing the contribution of several wavelengths (high energy $\gamma$-rays and X-rays). 

Continuing on the analysis of the comparison with the observational properties, we here match our theoretical simulations with the observed $\theta_{j}$. Also, the observed $\theta_{j}$ still carries many uncertainties due to the density medium approximation used in these calculations. To avoid the use of any theoretical assumption, here we consider the method of \cite{Pescalli2015MNRAS.447.1911P} 
to compare our simulated results with this estimate. In addition, following the discussion of  \cite{Lloyd2019MNRAS.488.5823L}, the observations of $\theta_{j}$ of LGRBs suggest that LGRB jets are narrower for those GRBs at higher redshift, thus we consider this conclusion in this paper to compute more realistic estimates of $\theta_{j}$. We have considered that SGRBs also undergo this evolution in our study since luminosities for SGRBs and LGRBs evolve similarly \citep{Dainotti2021ApJS..255...13D, PetrosianDainotti2024ApJ...963L..12P} and thus we have a similar expectation for the SGRBs. In the context of collimation by the stellar envelope, stars are generally characterised by lower metallicity at higher redshifts, which can collimate jets with larger efficiency. Although, in our calculation, we have fixed the values to high spin to allow the energy to be high to match the observations, it has been discussed the effect of BH spin on the jet collimation angle for a magnetically launched jet for a range of spin values \citep{Hurtado2024ApJ...967L...4H}. This means that, in principle, we can yet increase the spin to obtain a narrower jet, which would be consistent with the observed one even with a larger percentage of agreement.  
If we consider the evaluation of the $\theta_{j}$ method from \cite{Pescalli2015MNRAS.447.1911P}, we obtain $\theta_{j}$ = 6$^{\circ}$, and after we correct for redshift evolution, we reach 
a value of 10$^{\circ}$.

Our analysis is primarily based on 2D simulations, supplemented by a single 3D model (\texttt{HD-0.10-3D}). Although 3D simulations can capture inherently three-dimensional phenomena, such as magnetically arrested disk states and turbulence more accurately, performing extensive parameter studies in high-resolution 3D setups remains computationally demanding. MAD states in 2D may induce artificial turbulence in jets, potentially leading to a slight underestimation of $\theta_{jet}$. Furthermore, smoother accretion processes in 3D can yield marginally higher jet energetics compared to their 2D counterparts. Additionally, toroidal instabilities, which inherently require full three-dimensional treatment, may further influence jet dynamics and energetics. In this context, \cite{2022ApJ...933L...2G} conducted a set of 3D GRMHD simulations to investigate jet behaviors in short GRBs. They identified jet "wobbling," a toroidal instability causing periodic shifts in the jet direction. Such wobbling behavior can lead to intermittent emission peaks and variability in GRBs. This phenomenon is not observed in our current simulations, likely due to their limited evolution timescale. Despite these differences, our comprehensive 2D parameter study provides valuable insights into jet properties, capturing essential dynamics and yielding results that closely align with observational constraints for GRB 090510. Our findings thus set a viable foundation for future targeted high-resolution 3D investigations. 
In \cite{10.1093/mnrasl/slz012}, the simulated jet profile is analysed. Similarly, from Fig. \ref{figure:Gam_vs_th}, which gives information on $\Gamma_{\infty}$ and Fig. \ref{figure:op-angle-over-time}, which provides the time evolution of $\theta_{j}$ we can infer the jet structure. Specifically, Fig. \ref{figure:Gam_vs_th} gives the average picture of the averaged jet profile over time at radii of 3000 $R_{g}$. However, in Fig. \ref{figure:op-angle-over-time}, we present the evolution of $\theta_{j}$ from the beginning, binned in 0.05~s, until 0.6~s.

Finally, in our numerical model, we do not implement neutrino cooling, while one must remember that this component should be taken into account in a physical model. In \cite{2019MNRAS.482.3373F}, the neutrino cooling effects are considered, and the authors show that the neutrino-driven winds can act as an additional collimation mechanism of the SGRB jet; see also, e.g. \cite{2024arXiv240110094U}. To check to which extent the DE have an impact on the  $\theta_{j}$, we have added it in our calculation following the same profile of \cite{2022ApJ...933L...2G}, who added a specific profile of the medium behind the jet. They find the jet profile is changed, but they have a much extended computational grid.

\section{Summary and Conclusions}
\label{sec:summary and conclusions}

In this work, we have conducted detailed GRMHD simulations to investigate GRB 090510, a well-observed short GRB across multiple wavelengths \citep{DePasquale2010ApJ...709L.146D, 2016ApJ...831...22F, Dainotti2021ApJS..255...13D}. Through simulations, we evaluated a range of values for $\theta_{j}$ within our model. We have found angles of a few degrees, which is not unusual, as shown in the case of GRB 140903A \citep{Troja2016ApJ...827..102T}. Subsequently, we compared these values with observed values 
obtained through alternative methods, such as the one detailed in \cite{Pescalli2015MNRAS.447.1911P}. Furthermore, we have explored the scalability of this model in terms of energy and redshift ($z$). 
From our numerical simulations, we determined the structure of the jet across the cone so that we go beyond the frequently used, simplified top-hat scenario. We also provided a physical model of the jet's variable thermal and magnetic energy content and its ultimate Lorentz factor, $\Gamma$, estimations. Finally, we have linked the theoretically obtained jet properties with those determined for GRB 090510. While our primary analysis focused on GRB~090510, we also showed that the jet properties of several other short GRBs, including GRB~120804A with an opening angle independently constrained from afterglow observations, are consistent with our simulation models.

We highlight here that we reached the following takeaways points:
\begin{itemize}
\item The main drive for the jet collimation is represented by disk winds, with additional contributions from BH spin, magnetic field strength, and the proximity of the accretion disk. 

\item We show that the three models are equally representative of the property of the burst and exhibit an average opening angle of 9.2$^\circ$, 11.1$^\circ$, and 10$^\circ$. All these opening angles are compatible within 1 $\sigma$ with the observed $\theta_{j}$ of {$10.04^\circ \pm 1.29^\circ$} 
computed from the \cite{Pescalli2015MNRAS.447.1911P} relation after correction for the jet evolution, considering that both the Amati and the Ghirlanda relation carries no uncertainties. 

\item Similarly to the jet opening angle findings, we have observed compatibility of the variability values of 4.75 ms and 4.98 ms computed within two locations in the jet from our model and the wavelet analysis within 1 $\sigma$ as compared to the value of variability of $4.9^{+1.1}_{-0.9}$  discussed in \cite{2013MNRAS.432..857M}.

\item Although a single model cannot uniquely reproduce a specific GRB due to parameter degeneracies, our three models that match the observed $E_{\rm{GRB}}$ do so with efficiencies ranging from $\eta=9.5-13.7\%$. 

\item  The model with the smallest inner torus radii, placing it closest to the BH, produces a stable jet with the narrowest opening angle in all our models,  consistent with the angle inferred for GRB 090510. 

\item 3D simulation reveals a resolved MAD state that reduces variability and leads to smoother jet dynamics, enhancing our portrayal of realistic jet behavior in contrast to 2D models.

\end{itemize}

In conclusion, our work successfully achieves its primary goal of replicating properties of a target GRB. By detailing various models in Table \ref{tab:Sim_Models}, we also set the stage for more extensive studies of GRB jets by setting up initial parameter ranges for future simulations.  This integrated approach substantially advances our understanding of modelling high-energy GRBs.

\section*{Acknowledgements}

This work was supported by the grant 2019/35/B/ST9/04000 from the Polish National Science Center. 
A.J. was also partially supported by the NSC grant 2023/50/A/ST9/00527. 
We gratefully acknowledge Polish high-performance computing infrastructure PLGrid (HPC Center Cyfronet AGH) for providing computer facilities and support within computational grant no. 
PLG/2024/017347.
We acknowledge NAOJ's support for hosting J.S. and A.J. during their visits at NAOJ.  M.G.D. acknowledges the support of the JSPS Grant-in-Aid for Scientific Research (KAKENHI) (A), Grant Number JP25H00675. S.B. and M.G.D. acknowledge support from the Center for Computational Astrophysics at the National Astronomical Observatory of Japan, where 
part
of the numerical computations were performed using the Cray XC50 supercomputer. 
We also thank Tomoya Takiwaki and Hiroki Nagakura for their useful suggestions 
and Asaf Peer 
and Gerardo Urrutia for the useful discussions. 


\bibliographystyle{aa}
\bibliography{references.bib}
\end{document}